\documentclass[final,5p,times,twocolumn]{elsarticle}

\usepackage{amssymb}

\usepackage{lineno}
\usepackage{algorithmic}
\usepackage[]{algorithm2e}
\usepackage{amsmath}
\usepackage{multirow}
\usepackage{graphicx} 
\usepackage{caption} 
\usepackage{subcaption} 
\usepackage{placeins}
\usepackage{lipsum}

\usepackage{placeins}
\usepackage{graphicx} 
\usepackage{caption} 
\usepackage{subcaption} 
\usepackage{pgfplots}
\usepgfplotslibrary{dateplot}

\newcounter{groupcount}
\pgfplotsset{
    draw group line/.style n args={6}{
        after end axis/.append code={
            \setcounter{groupcount}{0}
            \pgfplotstableforeachcolumnelement{#1}\of#6\as\cell{%
                \def\temp{#2}
                \ifx\temp\cell
                    \ifnum\thegroupcount=0
                        \stepcounter{groupcount}
                        \pgfplotstablegetelem{\pgfplotstablerow}{X}\of#6
                        \coordinate [yshift=#4] (startgroup) at (axis cs:\pgfplotsretval,0);
                    \else
                        \pgfplotstablegetelem{\pgfplotstablerow}{X}\of#6
                        \coordinate [yshift=#4] (endgroup) at (axis cs:\pgfplotsretval,0);
                    \fi
                \else
                    \ifnum\thegroupcount=1
                        \setcounter{groupcount}{0}
                        \draw [
                            shorten >=-#5,
                            shorten <=-#5
                        ] (startgroup) -- node [anchor=base, yshift=0.5ex] {#3} (endgroup);
                    \fi
                \fi
            }
            \ifnum\thegroupcount=1
                        \setcounter{groupcount}{0}
                        \draw [
                            shorten >=-#5,
                            shorten <=-#5
                        ] (startgroup) -- node [anchor=base, yshift=0.5ex] {#3} (endgroup);
            \fi
        }
    }
}

\begin{document}

\begin{frontmatter}
\title{On the Efficient Design of Network Resilient to Electro-Magnetic Pulse Attack -- Elastic Optical Network Case Study }
\author{R\'o\.za Go\'scie\'n}

\address{Department of Systems and Computer Networks, \\ Faculty of Electronics, Wroclaw University of Science and Technology, Wroclaw, Poland, \\
e-mail: roza.goscien@pwr.edu.pl.}

\begin{abstract}

The telecommunication networks have become an indispensable part of our everyday life, providing support for such important areas as business, education, health care, finances, entertainment and social life. Alongside their continuous and uninterrupted operation is required while numerous new threats and attack scenarios emerge. The international security organisations warn against increasing likelihood of nuclear weapon or electro-magnetic pulse (EMP) attacks, which can be extremely harmful also for transport networks. On that background, we study efficient design of network resilient to EMP attack wherein the required protection level is provided by the application of multipath routing and military grade bunkers (advanced electro-magnetic radiation resilient approaches protecting whole network node) implementation. Formally, we define and study problem of bunkers location, routing and spectrum allocation (BLRSA) in elastic optical network (EON). In the problem objective we address two criteria -- network resilience (measured by the average lost flow per potential attack) and spectrum usage. For that problem we propose integer linear programming (ILP) model and two dedicated heuristics -- 1S-RSA and 2S-RSA. Then, we perform extensive numerical experiments divided into three parts: (\textit{i}) tuning of the proposed approaches, (\textit{ii}) comparison with reference methods, (\textit{iii}) realistic case study -- efficient EMP-resilient network design. In the case study we analyze benefits and costs of the proposed protection scheme. Moreover, we also analyze vulnerabilities of three realistic network topologies to EMP attacks and identify their critical nodes. The investigation proves high efficiency of the proposed approaches and shows that they allow to save up to 90\% of traffic lost in the case of no protection against these types of attacks.

\end{abstract}


\end{frontmatter}

\section{Introduction}

The beginnings of telecommunication networks can be found in 1960s when the ARPANET (advanced research project group network) was invented. It was a~simple packet-forwarding network supporting only a~few basic services. Since that, the networks have undertaken enormous changes whose were triggered by the changing and increasing requirements of the network users. One of the most important factors affecting the network development is growing number of network users and connected devices. Cisco company estimates that by the year 2023 there will be 5.3 billion global Internet users. On top of that, they assess that there will be 3.6 global devices and connections per capita~\cite{Cisco_air}. Moreover, we also observe that users are increasingly interested in bandwidth intensive services such as cloud computing, content distribution or online audio/video streaming~\cite{Cisco_air}. As a~result of these trends, the overall traffic volume in the networks rises rapidly, which is especially noticeable in core networks. The solutions and techniques that are used nowadays are expected to be not efficient enough for near future networks. It, therefore, triggers urgent need for some improvements of currently applied approaches or completely new and more efficient technologies. One of the most promising solutions proposed for optical core networks is architecture of elastic optical network (EON) \cite{Jinno_commag10, Christodoulopoulos_lightwave11}. EON provides superior spectrum utilization (compared to the previously deployed and still popular wavelength division multiplexing, WDM), as it operates within flexible frequency grids and supports for advance modulation and transmission techniques \cite{Goscien_network15}. In more detail, EON introduces a~new spectrum provisioning manner -- the entire available spectrum width is divided into narrow and same-sized segments called \textit{slices}. Then, communication channels, which can be tailored to the incoming demands' bitrates, are created by grouping a~number of adjacent slices. 

Increasing popularity of telecommunication networks in our life and their support for variety of our activities (including business, education, medical care, financial, social networking and entertainment), implies a~requirement of provisioning their uninterrupted operation. Networks rely on the imperfect devices (which are always prone to failures) and, as a~crucial infrastructure for our society, may be a~target of an attack. Hence, it is important to design and optimize them in such a~way to minimize probability of their failures and mitigate inevitable failures'/attacks' consequences \cite{Rak_networks20, Goscien_network15}. It is especially important to protect networks against potential attacks, considering they are always planned to be the most harmful to network as it is possible \cite{SkorinKapov_commag16, Furdek_rndm16}. Depending on the size of attack-affected areas (and number of affected elements), we can distinguish small- and large-scale attacks (or generally failures). The former ones are trivial to be addressed by means of simple protection methods such as dedicated or shared path protection. The latter ones are more challenging and therefore need advanced and sophisticated methods. 

Following the analysis of international security organizations the nuclear weapon and the electro-magnetic pulse (EMP) are becoming dangerous and likely to be performed threats \cite{nuclear_summit_16}. Therefore, they should be addressed in various security and protection strategies, including transport network protection plans. A~nuclear weapon attack is always targeted at a~specific location. Due to the radioactive mass explosion, a~specific area is then destroyed (i.e., attack zone). The size of that area depends mostly on the radioactive element mass and the explosion height. However, the attack is also followed by the EMP -- a~short burst of electro-magnetic energy. As a~result, an excessive electro-magnetic (EM) radiation propagates over long distance through the attack zone and might affect various electronic and electric elements/devices. The range of the radiation is difficult to be determined, since it depends on plenty of various factors (i.e., initial pulse energy, explosion height above the ground, latitude and longitude of attacked target, shape terrain and buildings, weather conditions, etc.). Furthermore, realistic case studies cannot be performed \cite{metatech_2010}. Moreover, it is nearly impossible to predict the behaviour of an electronic device under such an excessive EM radiation level. Elements and modules of devices are tested in order to determine their vulnerability to the radiations. The results strongly depend on their construction (which may vary between the vendors), applied materials and elements'/modules' assembly. It is worth-mentioning that it is possible to construct a~device or system that is resistant to EM radiation at a~specific level \cite{Wang_sege13,Brauer_isec09,Palisek_isec11}. The realization relies on the military grade bunkers or protection approaches, which base on the shielding of equipment as well as entire rooms/building with protected infrastructure. For the sake of simplicity, we refer to these protection approaches as~\textit{bunkers}. The technology is limited when it comes to protecting network efficiently against destruction caused by the nuclear weapon explosion. Yet, we might design and optimize network to be resilient against the following EMP and excessive EM radiation. First, considering to protect some network nodes (precisely -- the equipment located in these nodes) by placing the equipment in the military grade bunkers or protection. But it significantly increases network cost, hence, the available budget will definitely limit number of bunkers possible to be implemented. Second, we can design dedicated path-based protection strategies for demands to improve their resilience to nuclear/EMP attacks. 

In this paper, we focus on the efficient design of optical network, which simultaneously addresses issues of spectrum usage and protection against EMP attacks. To provide high network resilience, we combine benefits of bunkers' implementation and multipath routing. As a~case study optical network technology, we use promising elastic optical network. We define and analyze problem of bunkers location, routing and spectrum allocation (BLRSA) in elastic optical networks. Please note that BLRSA combines three optimization tasks: (\textit{i})~bunkers location (BL), (\textit{ii})~routing assignment (RA) and (\textit{iii})~spectrum assignment (SA). We formulate that problem using integer linear programming (ILP) technique and, since it is very challenging, propose a~bunch of heuristic solution methods which decompose BLRSA into at least two subsequent phases of BL and RSA. To solve BL task, we propose five different policies, wherein two of them are completely novel. To solve RSA, we propose two dedicated heuristics -- 1S-RSA and 2S-RSA, Further, we adapt two reference methods given in the literature for similar problem versions. Next, we perform extensive numerical experiments focused on three aspects: (\textit{i})~tuning of the proposed approaches, (\textit{ii})~comparison of all solution methods and determination of best one, (\textit{iii})~case study -- efficient EON design considering spectrum usage and protection against EMP attacks. The case study is a~vast part of experiments, in which we determine efficiency and costs of the proposed protection scheme. We also analyze vulnerabilities to EMP attacks for three realistic network topologies.  

The rest of paper is organized as follows. In Section~\ref{sec:related_works} we review the related works. In Section~\ref{sec:blrsa-model} we define optimization problem while in Section~\ref{sec:algorithms} we present solution algorithms. Next, in Section~\ref{sec:experiments} we report and analyze results of numerical experiments. In the last Section~\ref{sec:conclusions} we conclude whole paper. 

\section{Related works}\label{sec:related_works}

As a~highly promising technology for optical networks, EONs have been extensively studied in the literature including their resiliency provisioning \cite{Shen_pnc16, Hai_comcom19, Goscien_comcom18}. In terms of small-scale failures (i.e., a~single or several elements failure), mainly dedicated/shared path-based and p-cycle methods were applied regardless of the failure reason. Concurrently, large scale failures can be classified as natural disasters or man-made attacks, depending on the failure cause \cite{Furdek_rndm16}. In the case of natural disaster, it is possible to predict and precisely model the disaster scenario (such as earthquake, flood, hurricane, etc) and then use these approaches in order to design efficient protection/restoration strategies \cite{Pasic_rndm18, Pasic_rndm19, Agrawal_jlt19, Valentini_rndm19, Agarwal_trannet13}. Note that most commonly addressed natural disaster scenario in transport networks is an earthquake \cite{Agrawal_jlt19, Valentini_rndm19}. On the contrary to natural disasters, man-made attacks are always pre-planned in order to be most harmful, so they cannot be modeled in the same way as the natural disasters. In order to improve network resilience to such attacks it is necessary to identify potential attack scenarios, network vulnerabilities (including critical links/nodes) and then use these information to design/optimize efficient protection/restoration mechanisms \cite{Agarwal_trannet13, Neumayer_trannet11, Barbosa_rndm18}. It is also worth-mentioning, that majority of efficient protection/restoration algorithms makes use of identification of so called shared risk groups (SRGs). I.e., sets of nodes and links which might be affected simultaneously by a~single disaster event \cite{Nedic_rndm18, Habib_jlt12, Tapolcai_infocom17}. The literature review shows also high efficiency of application of geodiverse routing, i.e., multipath demands allocation with paths that are geographically most distant \cite{Nedic_rndm18, Sousa_drcn17}.

Concurrently, various aspects of EMPs and excessive EM radiation were investigated in the literature, where two most important of them are: modeling of EMP and corresponding EM radiation and assessment \cite{metatech_2010,Wang_sege13} of its harmful effects \cite{Brauer_isec09,Kohlberg_apsursi11,Palisek_isec11,Xianguo_ceem15}. The research related to modeling of EMP and its corresponding EM radiation aims mainly on the determination of the size and shape of the effected area (from the point source of radiation) and estimation of the radiation level in any point of the affected area. The related models are complex, while their formulas depend on plethora of factors such as characteristics of the radiation source, its exact location (longitude and attitude), characteristics of the propagation area (i.e., its topography, density, height and the structure of buildings), weather conditions, etc. Then, the research related to assessment of EM radiation harmful effects studies behaviour of various electronic elements, modules, devices and systems under some specific excessive radiation levels. They also provide guidelines on how to design EM radiation-resilient electronic equipment and how to protect existing  components. The protection strategies might be costly and therefore are applied only for mission-critical equipment~\cite{Brauer_isec09}. 

The research related to resilient transport networks lacks studies which directly address nuclear weapon/EMP attacks considering their specific characteristics (such as potential destructive and jamming ranges) and possibility to protect some network elements/nodes by means of bunkers. More precisely, the research focuses on the protection against large scale disasters, where an attack affects large geographically correlated areas and complete protection of selected devices/equipment is mostly not possible (note that in the case of EMP attack nodes can be protected by the bunkers). The only paper that directly addresses EMP attack is \cite{Massimo_icc17}, however, it focuses on the post-disaster data evacuation from isolated data centers through LEO satellite. Therefore, this paper fills the literature gaps and study post nuclear attack EMP-resilient transport network design considering both destructive and jamming attack ranges and possibility to protect network nodes with bunkers.  

\section{BLRSA problem formulation}\label{sec:blrsa-model}

In this section, we define formally considered problem of bunkers location, routing and spectrum allocation (BLRSA). 

EON is modeled as a directed graph $G=(V,E)$, where $v \in V$ are network nodes and $e \in E$ are directed physical links. Each physical link \textit{e} is given by its source node $o(e)$, destination node $t(e)$ and length in kilometres $l_e$. For each node \textit{v}, we identify sets of outgoing $E^+(V)$ and incoming $E^-(V)$ links. The spectrum width available on each physical link is divided into frequency slices $s \in S$. By grouping adjacent slices we can create frequency channels $c \in C$. Each channel \textit{c} is characterized by the first slice index and number of involved slices $\beta_c$. 

Based on the preliminary studies, we are given with a~set of identified potential nuclear attack scenarios $a \in A$. Each one targeted at a~specific network node and, depending on the explosion height and radioactive mass, characterized by a~destructive and jamming ranges. Please note that the destructive range is a~direct result of the radioactive mass explosion while jamming range is a~repercussion of the following EMP. Since optical network considered, it is assumed that links are resistant to EM radiation of excessive level and EMP might affect only devices located in network nodes. In turn, we have binary constants $\alpha_{va}$ and $\gamma_{ea}$ which inform if node~\textit{v} is, accordingly, in the destructive or jamming range of a~potential attack \textit{a}. It is assumed that all nodes located in the destructive region are destroyed and thus do not operate at all while nodes located only in the jamming region stopped to operate unless they are protected by a~bunker. Please note that network link works properly only if both of its ends (i.e., source and destination node) operate correctly. Since bunkers are expensive, we can locate them only in at most $|B|$ nodes while the decision where to place bunkers $b \in B$ is the~part of the optimization task.  

A~set of point-to-point traffic demands $d \in D$ is to be realized. Each demand \textit{d} is characterized by its source node $s_d$, destination node $t_d$ and bitrate $h_d$ given in Gbps. Since EON technology is considered, we have to tackle the problem of routing and spectrum assignment (RSA). In order to increase network resilience against attacks, multipath routing is applied. In turn, each demand has to be assigned with $|P|$ different routing paths $p \in P_d$. Then, each of these paths has to be assigned with a~frequency channel tailored to the demand bitrate and path length. By these means, each demand has to be finally assigned with $|P|$ different light-paths $l \in L_d$. 

To realize data transmission, four modulations $m \in M$ are available -- 32-QAM, 16-QAM, QPSK, BPSK. For a~particular traffic demand \textit{d} and its candidate routing path~\textit{p}, we always select the most spectrally efficient modulation format which satisfies transmission on the path \textit{p} (according to its length). For the purpose of modeling, let \textit{m}=1 denotes the most spectrally efficient modulation format (32-QAM) and \textit{m}=4 refers to the least efficient one (BPSK). Based on the model reported in Table~\ref{tab:modulations}, we calculate two sets of constants: (\textit{i})~$a_m, m\in M$ as a~lower bound of the distance range supported by the modulation~\textit{m}, (\textit{ii})~$n_{dm}, d \in D, m \in M$ as a~number of slices required to realize demand \textit{d} using modulation \textit{m}. Then, we define a~set of constants $h_{dm}$ indicating the number of additional slices required to realize demand \textit{d} using modulation format \textit{m} instead of \textit{m}-1. Hence, $h_{d1} = n_{d1}$ and $h_{dm} = n_{dm} - n_{d(m-1)}, m>1, m \in M$.

\begin{table}[htb]
    \centering
    \caption{Supported bit-rate and transmission distance for a~transponder operating within 37.5~GHz spectrum \cite{Khodashenas_lighttech16}}
    \label{tab:modulations}
    \begin{footnotesize}
\begin{tabular}{|c|c|c|c|c|}\hline
        
        & \textbf{BPSK} & \textbf{QPSK} &\textbf{ 32-QAM} &\textbf{ 16-QAM} \\ \hline
       \textbf{ supported bitrate} [Gbps] & 50 & 100 & 150 & 200 \\ \hline
        \textbf{transmission reach} [km] & 6300 & 3500 & 1200 & 600 \\ \hline
    
    \end{tabular}
    \end{footnotesize}
\end{table} 

Please also note that having information regarding $n_{dm}$, for each demand \textit{d} and its routing path \textit{p} (the path length determines applied modulation format), we can reduce the set of candidate channels to $c \in C_{dp} \in C$ by considering only channels grouping $n_{dm}$ slices.

The aim of the optimization problem is to locate $|B|$ bunkers in the network and assign each traffic demand with a~set of $|P|$ light-paths in such way it minimizes the spectrum usage and network vulnerability to any of the identified potential attack scenarios. Since both criteria cannot be optimized simultaneously, we introduce a~weighted objective function (see formula~(\ref{node_design_obj})) with coefficients $c_{spec}$ and $c_{res}$ establishing priorities between two optimization criteria. Note that $0 \leq c_{spec}, c_{res} \leq 1$ and $c_{spec}+c_{res}=1$. The criterion related to the spectrum usage is defined as the~maximum obtained spectrum usage in the network (required to realize all demands) divided by the $MAX\_SPEC$. Then, criterion related to the network resilience is defined as the~average lost flow per attack divided by the $MAX\_LOSS$. The values of $MAX\_SPEC$ and $MAX\_LOSS$ are set arbitrary based on characteristics of testing scenarios. Their values are selected in such way that satisfies the condition $c_{spec}, c_{res} \leq 1$ and make values of both criterions similar according to order of the magnitude.

The problem constraints are given by the formulas (\ref{onestep_flow_conservation})--(\ref{onestep_slice_capacity}). Equation~(\ref{onestep_flow_conservation}) is a~flow conservation condition. Constraint~(\ref{onestep_bunkers_limit}) assures that at most $|B|$ bunkers are located. Formulas (\ref{onestep_destructive_range}) and (\ref{onestep_jamming_range}) define variable $q_{va}$, which informs if node~\textit{v} operates properly in the case of attack \textit{a}. Formula (\ref{onestep_destructive_range}) checks if node \textit{v} is in the destructive range of attack \textit{a} while formula (\ref{onestep_jamming_range}) takes into account jamming range and potential protection by means of a~bunker. Next, variable $z_{ea}$ is defined, which informs whether link~\textit{e} works properly in the case of attack~\textit{a}. Recall that link is available when both source node (eq.~(\ref{onestep_available_src_node})) and destination node (eq.~(\ref{onestep_available_dst_node})) operate properly. Formula~(\ref{onestep_path_availability_in_a}) verifies availability of paths selected for demands in case of a~specific attack scenario. Then, equation~(\ref{onestep_demand_realization_in_a}) checks if demand~\textit{d} is realized in the case of attack \textit{a} taking into account selected paths and their availability. Constraint~(\ref{onestep_channel_selection}) assures that exactly one channel is selected for each demand on each selected path. Formula~(\ref{onestep_pathlength}) calculates the length of each selected routing path. Constraint~(\ref{onestep_modulation_selection}) is the modulation selection for each demand and routing path. Equation~(\ref{onestep_channel_size}) assures that channel with appropriate size is selected for each demand and routing path. Eventually, constraints~(\ref{onestep_qpdes}) and (\ref{onestep_slice_capacity}) control the spectrum usage and non-overlapping.  

\noindent\textbf{Sets and indices:}\\
\noindent \begin{tabular}{@{} p{0.13\linewidth}p{0.82\linewidth} @{}}

		$ v \in V$ & network nodes \\
		$ e \in E$ & network links \\
		$E^+(v)$ & set of links that originate in node \textit{v} \\
        $E^-(v)$ & set of links that terminate in node \textit{v} \\
        $d \in D$ & traffic demands to be realized \\
		$p \in P$ & routing paths required for each demand \\
		$ s \in S$ & frequency slices available on each network link \\
		$c \in C$ & candidate frequency channels created for $|S|$ slices \\
        $ m \in M$ & available modulation formats \\
        $a \in A$ & identified potential attack scenarios \\
        
\end{tabular}

\noindent\textbf{Constants:}\\
\noindent \begin{tabular}{@{} p{0.13\linewidth}p{0.82\linewidth} @{}}

    	$h_d$ & bitrate (in Gbps) of demand \textit{d} \\
        $s_d$ & source node of demand \textit{d} \\
        $t_d$ & destination node of demand \textit{d} \\
        $l_e$ & length (in kilometers) of physical link \textit{e} \\
        $s(e)$ & source node of physical link \textit{e} \\
        $t(e)$ & destination node of physical link \textit{e} \\
        $\beta_{c}$ & size (number of involved slices) of channel \textit{c} (excluding a~guard-band) \\ 
        $\delta_{cs} $ & =1, if channel \textit{c} involves slice \textit{s}; 0, otherwise \\
    
        $n_{dm}$ & number of slices required to realize demand \textit{d} using modulation \textit{m} \\
        
        

\end{tabular}

\noindent \begin{tabular}{@{} p{0.22\linewidth}p{0.68\linewidth} @{}}
        
        $\alpha_{va}$ & =1, if node \textit{v} is in the destructive range of attack~\textit{a}; 0, otherwise \\
        $\gamma_{va}$ & =1, if node \textit{v} is in the jamming range of attack~\textit{a}; 0, otherwise \\
        
        $h_{dm}$ & number of additional slices required for demand~\textit{d} if modulation~\textit{m} is applied instead of~\textit{m}-1 \\
        $r_{m}$ & lower bound of the distance range supported by modulation format \textit{m} \\
        $c_{spec}$ & weight of the spectrum usage in the objective function. Note $0 \leq c_{spec} \leq 1$ \\
        $c_{res}$ & weight of the network resiliency in the objective function. Note $0 \leq c_{res} \leq 1$ \\
        $L$ & large number \\
        $MAX\_SPEC$ & constant defining optimization criterion related to spectrum usage  \\
    $MAX\_RES$ & constant defining optimization criterion related to network resilience  \\

\end{tabular}

\noindent\textbf{Variables:}\\

\noindent \begin{tabular}{@{}p{0.1\linewidth}p{0.85\linewidth} @{}}
	$ b_{v}$ & =1, if node \textit{v} is protected by a~bunker; 0, otherwise (binary) \\
	
	$f_{dpe}$ & =1, if demand \textit{d} uses link \textit{e} on path \textit{p}; 0, otherwise (binary) \\
	$w_{dpc}$ & =1, if demand \textit{d} uses channel \textit{c} on path \textit{p}; 0, otherwise (binary) \\
	$u_{dpa}$ & =1, if path \textit{p} selected for demand \textit{d} is available in the case of attack \textit{a}; 0, otherwise (binary) \\
	$x_{da}$ & =1, if demand \textit{d} is realized in the case of attack \textit{a}; 0, otherwise (binary) \\
	$n_{dpm}$ & =1, if any modulation format $i < m$ cannot be applied for demand \textit{d} on its path \textit{p}; 0, otherwise
(binary) \\
	$v_{dp}$ & length for path \textit{p} selected for demand \textit{d} (integer) \\
	
	$ y_{dpes}$ & =1, if path \textit{p} selected for demand \textit{d} uses slice \textit{s} on link \textit{e}; 0, otherwise (binary) \\
	$ y_{s}$ & =1, if slice \textit{s} is used in the network; 0, otherwise (binary) \\
	
	$q_{va}$ & =1, if node \textit{v} is available in the case of attack \textit{a}; 0, otherwise (binary) \\
	$z_{ea}$ & =1, if link \textit{e} is available in the case of attack \textit{a}; 0, otherwise (binary) \\

\end{tabular}

\noindent
\textbf{objective}
\begin{small}
\begin{equation}\label{node_design_obj}
\min (c_{spec} \cdot \frac{\sum_{s \in S} y_s}{MAX\_SPEC}  + c_{res} \cdot \frac{\sum_{a \in A} \sum_{d \in D} (1 - x_{da}) \cdot h_d}{MAX\_LOSS} )
\end{equation}
\end{small}

\noindent
\textbf{subject to}
\begin{small}
\begin{align}\label{onestep_flow_conservation}
\sum_{e \in E^+(v)} f_{dpe} - \sum_{e \in E^-(v)} f_{dpe} = 
\begin{cases}
    1 & \text{ if } v = s_d \\
    -1 & \text{ if } v = t_d \\
    0 & \text{ if } v \neq s_d, t_d \\
\end{cases} \nonumber \\
d \in D, p \in P, v \in V.
\end{align}

\begin{equation}\label{onestep_bunkers_limit}
\sum_{v \in V} b_v \leq B.
\end{equation}

\begin{equation}\label{onestep_destructive_range}
q_{va} \leq 1 - \alpha_{va}, a \in A, v \in V.
\end{equation}

\begin{equation}\label{onestep_jamming_range}
q_{va} \leq 1 - \gamma_{va} + b_v, a \in A, v \in V.
\end{equation}

\begin{equation}\label{onestep_available_src_node}
z_{ea} \leq q_{s(e)a}, e \in E, a \in A.
\end{equation}

\begin{equation}\label{onestep_available_dst_node}
z_{ea} \leq q_{t(e)a}, e \in E, a \in A.
\end{equation}

\begin{equation}\label{onestep_path_availability_in_a}
u_{dpa} \leq z_{ea} - f_{dpe} + 1, d \in D, p \in P, a \in A, e \in E.
\end{equation}

\begin{equation}\label{onestep_demand_realization_in_a}
x_{da} \leq \sum_{p \in P} u_{dpa}, d \in D, a \in A.
\end{equation}

\begin{equation}\label{onestep_channel_selection}
\sum_{c \in C} w_{dpc} = 1, d \in D, p \in P.
\end{equation}

\begin{equation}\label{onestep_pathlength}
\sum_{e \in E} f_{dpe} \cdot l_e \leq v_{dp}, d \in D, p \in P.
\end{equation}

\begin{equation}\label{onestep_modulation_selection}
v_{dp} - r_m \leq L \cdot n_{dpm}, m \in M, d \in D, p \in P.
\end{equation}

\begin{equation}\label{onestep_channel_size}
\sum_{m \in M} n_{dpm} \cdot h_{dm} \leq \sum_{c \in C} w_{dpc} \cdot \beta_c, d \in D, p \in P.
\end{equation}

\begin{equation}\label{onestep_qpdes}
f_{dpe} + \sum_{c \in C} w_{dpc} \cdot \delta_{cs} - 1 \leq y_{dpes}, d \in D, p \in P, s \in S, e \in E.
\end{equation}

\begin{equation}\label{onestep_slice_capacity}
\sum_{e \in E}\sum_{d \in D} \sum_{p \in P} y_{dpes} \leq y_s, s \in S.
\end{equation}

\end{small}

\section{Heuristic algorithms}\label{sec:algorithms}

In this section, we present heuristic algorithms proposed to solve the BLRSA. Each algorithm solves BLRSA problem by dividing it into at least two subsequent tasks: (\textit{i})~bunkers' location, (\textit{ii})~routing and spectrum assignment for a~given set of already placed bunkers. For a~bunkers' location, we propose five different policies, which are discussed in Section~\ref{sec:bunkers_policies}. To solve RSA problem for a~given set of already placed bunkers, we propose two completely novel algorithms called 1S-RSA (see  Section~\ref{sec:1s-rsa}) and 2S-RSA (see  Section~\ref{sec:2s-rsa}). 1S-RSA (1-step routing and spectrum assignment) method solves simultaneously problems of routing and spectrum assignment. 2S-RSA (2-steps routing and spectrum allocation) approach divides it into two subsequent tasks. In the first step it solves routing problem and then, in the second step, it assigns spectrum resources. In order to obtain reference methods, we also adapt two popular RSA algorithms for the problem formulation considered in this paper (see  Section~\ref{sec:ref-algs}). Reference methods are called: first-fit routing and spectrum assignment (FF-RSA) and link-disjoint routing and spectrum assignment (LD-RSA). 

\subsection{Bunkers' location policies}\label{sec:bunkers_policies}

For the bunkers' location task, we propose five different policies -- AvgNeighbour, MinNeighbour, NodalDegree, Adaptive/Avg and Adaptive/Max. 

Three first policies are based on the characteristics of the considered network topology. The main idea behind AvgNeighbour and MinNeighbour policies lays in the surmise that nodes that have close (in terms of geographical distance) neighbours will be potentially more frequently used as intermediate nodes for transmissions, and therefore might become a~target of an attack. In turn, AvgNeighbour policy calculates an average distance to a~neighbour (a~directly connected node) for each network node. Then, it places bunkers in the nodes for which the obtained criterion is the smallest. MinNeighbour runs similarly, however, it calculates a~minimum distance to a~neighbour for each network node. The premise behind NodelDegree is fact that also nodes with high nodal degree might be selected as intermediate nodes of many selected paths and in turn -- targets of attacks. Therefore, NodalDegree policy places bunkers in nodes with highest nodal degree. 

Adaptive policy is a~completely novel proposal which bases on the analysis of the network topology, already placed bunkers and identified potential attack scenarios. We consider two versions of that policy -- Adaptive/Avg and Adaptive/Max. The idea behind these methods lays in the construction of the network vulnerability matrix $VUL\_M$, which estimates potential failure risk of each network link. The policy works in $|B|$ iterations and in each iteration one bunker location is selected. In the beginning of each iteration, a~network vulnerability matrix is constructed. This is a~$|V| \cdot |V|$ matrix, where element in \textit{i}-th row and \textit{j}-th column denotes vulnerability metric of the link connecting nodes \textit{i} and \textit{j}. In the considered research, a~link vulnerability metric is equal to the number of potential attacks (from the set of the identified attacks) which lead to the link unavailability according the attacks' ranges and already located bunkers. Please note that in the \textit{i}-th iteration, (\textit{i}-1) bunkers are already located in network nodes. Having network vulnerability matrix, we calculate a~vulnerability metric for each network node as an average (for Adaptive/Avg) or maximum (for Adaptive/Max) link vulnerability metric among all incoming and outcoming links. Eventually, the first node with highest vulnerability metric is selected for a~bunker location. 

\subsection{1S-RSA}\label{sec:1s-rsa}

The idea of 1S-RSA is presented in Algorithm~\ref{alg:1s-rsa}. In the beginning, the method calculates network vulnerability matrix $VUL\_M$ (Alg.~\ref{alg:1s-rsa}, line~\ref{alg:1s-find-matrix}) for a~given topology and already located bunkers. Then, demands are sorted in decreasing bitrate (Alg.~\ref{alg:1s-rsa}, line~\ref{alg:1s-sort}) and handled one by one. For each demand, a~special function \textit{FindNextLightpath()} is run $|P|$ times (Alg.~\ref{alg:1s-rsa}, line~\ref{alg:1s-findlightpath}) to find required light-paths wherein in each iteration $i > 1$ all already selected light-paths are taken into account while finding next ones. The process of \textit{FindNextLightpath()} is presented in Algorithm~\ref{alg:findnextlightpathpath}. In each iteration, the method calculates $\lambda$ candidate routing paths. It first tries to find paths link-disjoint with already selected light-paths (Alg.~\ref{alg:findnextlightpathpath}, line~\ref{alg:fnlp_find_sld_paths}). If there are no such paths or the number of these paths is less than $\lambda$, then a~modified network vulnerability matrix $VUL\_M^\prime$ is calculated  (Alg.~\ref{alg:findnextlightpathpath}, line~\ref{alg:fnlp_find_modified_matrix}). Each element of that matrix (i.e., element in the \textit{i}-th row and \textit{j}-th column) is equal to the corresponding element of the network vulnerability matrix multiplied by the link (\textit{i},\textit{j}) popularity factor. For a~particular network link, we define its popularity factor as a~number of previously selected light-paths which use that link increased by 1.0.  Then, the set of candidate routing paths is filled to $\lambda$ elements by finding shortest paths according to metrics from modified network vulnerability matrix (Alg.~\ref{alg:findnextlightpathpath}, line~\ref{alg:fnlp_find_rest_paths}). Next, $\lambda$ candidate light-paths are created by finding first-fit channels (i.e., a~channel using lowest possible slices) for each of the selected routing paths (Alg.~\ref{alg:findnextlightpathpath}, line~\ref{alg:fnlp_find_channel}). To select best of candidate light-paths, a~special metric $\theta(l)$ is calculated according to formula~(\ref{eq:lightpath_cost}), which should be minimized (Alg.~\ref{alg:findnextlightpathpath}, line~\ref{alg:fnlp_find_best_lightpath}). 

\begin{equation}\label{eq:lightpath_cost}
\theta(l) = c_{spec} \cdot \frac{S\_MAX(l)}{|S|} + c_{res} \cdot \frac{LEN(l)}{LEN\_MAX} 
\end{equation}

\noindent Where:	

\noindent \begin{tabular}{@{}p{0.15\linewidth}p{0.8\linewidth} @{}}
$\theta(l)$ & cost of candidate light-path \textit{l}  \\
$c_{spec}, c_{res}$ & weights of the spectrum usage and network resilience in objective function, $0.0 \leq c_{spec}, c_{res} \leq 1.0, c_{spec}+c_{res}=1.0$  \\
S\_MAX(\textit{l}) & index of the last allocated slice on light-path \textit{l} \\
$|S|$ & number of slices available in the network \\
\end{tabular}

\noindent \begin{tabular}{@{}p{0.15\linewidth}p{0.8\linewidth} @{}}
LEN(\textit{l}) & length of light-path \textit{l} according to metrics from modified network vulnerability matrix  \\
LEN\_MAX & maximum length of a~light-path in the network, LEN\_MAX = $|A| \cdot |E|$  \\
\end{tabular}

\begin{algorithm}
\caption{1S-RSA algorithm}\label{alg:1s-rsa}

\begin{small}

\begin{flushleft}
        \textbf{INPUT:} network topology, potential attack scenarios, bunkers' locations, demands $d \in D$ to be allocated, parameters: $\lambda$, $c_{spec}$, $c_{res}$ and $|P|$ \\ 
        \textbf{OUTPUT:} allocation rules for all demands
\end{flushleft}

\begin{algorithmic}[1]

\STATE  calculate $VUL\_M$ \label{alg:2s-rsa-sort}\label{alg:1s-find-matrix}
\STATE  sort $d \in D$ in decreasing $h_d$ \label{alg:1s-sort}
\FOR{\textbf{each} $d \in D$}
\STATE $L_d \gets \emptyset $
\ENDFOR

\FOR{\textbf{each} $d \in D$} \label{alg:1s-rsa-start}
\FOR{$p \in P$}
\STATE $l \gets$ \textit{FindNextLightPath}(..., $VUL\_M$, $L_d$, $\lambda$, $c_{spec}$, $c_{res}$)\label{alg:1s-findlightpath}
\STATE $L_d = L_d \cup l$
\STATE allocate light-path \textit{l} for demand \textit{d}
\ENDFOR
\ENDFOR \label{alg:1s-rsa-end}

\end{algorithmic}
\end{small}
\end{algorithm}

\begin{algorithm}
\caption{\textit{FindNextLightPath()} function}\label{alg:findnextlightpathpath}

\begin{small}

\begin{flushleft}
        \textbf{INPUT:} network topology, potential attack scenarios, bunkers' locations, network vulnerability matrix $VUL\_M$, demand \textit{d}, already selected light-paths $l \in L_d$ for demand \textit{d}, parameters: $\lambda$, $c_{spec}$ and $c_{res}$  \\ 
        \textbf{OUTPUT:} selected light-path for demand \textit{d}
\end{flushleft}

\begin{algorithmic}[1]

\STATE $P \gets$ $Find\_SLD\_Paths$(..., $L_d$, $VUL\_M$, $\lambda$)\label{alg:fnlp_find_sld_paths}
\IF{$|P| \neq \lambda$}
\STATE $VUL\_M^\prime$ $\gets$ $FindModified\_VUL\_M$($L_d$, $VUL\_M$)\label{alg:fnlp_find_modified_matrix}
\STATE $P = P \cup $ \textit{FindShortestPaths}(...,$VUL\_M^\prime$, $\lambda-|P|$)\label{alg:fnlp_find_rest_paths}
\ENDIF

\STATE $L^\prime \gets \emptyset$
\FOR{$p \in P$}
\STATE $l \gets (p,$ \textit{FindChannel}(...,p)) \label{alg:fnlp_find_channel}
\STATE  $L^\prime \gets L^\prime \cup l$
\ENDFOR

\RETURN $l^\prime \gets \min_{l \in L^\prime} \theta(l)$ \label{alg:fnlp_find_best_lightpath}

\end{algorithmic}
\end{small}
\end{algorithm}

1S-RSA algorithm has one tuning parameter (i.e., $\lambda$) while values of $c_{spec}$ and $c_{res}$ characterize considered testing scenario. Please note that 1S-RSA solves problems of routing and spectrum allocation simultaneously, however, due to the existence of metric (\ref{eq:lightpath_cost}) it might prioritize two objective criteria (spectrum usage and resiliency) in a~different ways. 

\subsection{2S-RSA}\label{sec:2s-rsa}

The solving process of 2S-RSA is divided into two subsequent tasks -- routing assignment and spectrum assignment. The main method idea is presented in Algorithm~\ref{alg:2s-rsa}.

In the beginning, 2S-RSA constructs a~network vulnerability matrix $VUL\_M$ (Alg.~\ref{alg:2s-rsa}, line~\ref{alg:2s-find-matrix}) and sorts demands in decreasing bitrate (Alg.~\ref{alg:2s-rsa}, line~\ref{alg:2s-sort}). After that, the method moves to the routing assignment phase (Alg.~\ref{alg:2s-rsa}, lines~\ref{alg:2s-ra-start}--\ref{alg:2s-ra-end}) in which the sorted demands are handled one by one. For each demand, a~special function \textit{FindNextPath()} is run $|P|$ times to find required routing paths (Alg.~\ref{alg:2s-rsa}, line~\ref{alg:2s-findpath}). The idea of that method is presented in Algorithm~\ref{alg:findnextpath}. Initially, the procedure tries to find the shortest path (according to metrics from network vulnerability matrix) which is simultaneously link-disjoint with already selected paths (Alg.\ref{alg:findnextpath}, line~\ref{alg:fnp_find_sld_path}). If such a~path does not exist, the method calculates a~modified network vulnerability matrix $VUL\_M^\prime$ (Alg.\ref{alg:findnextpath}, line~\ref{alg:fnp_find_modified_matrix}). Then, the shortest path is selected according to the metrics from modified vulnerability matrix (Alg.\ref{alg:findnextpath}, line~\ref{alg:fnp_find_spath_in_modified}). In the next step, 2S-RSA moves to the spectrum assignment phase (Alg.\ref{alg:2s-rsa}, lines~\ref{alg:2s-sa-start}--\ref{alg:2s-sa-end}), which handles demands in the same order as previously. For each of the selected paths $p \in P_d$, 2S-RSA finds first free spectrum channel (i.e., channel characterized by the minimum last slice index) and allocates demands (Alg.\ref{alg:2s-rsa}, lines~\ref{alg:2s-findchannel}-- \ref{alg:2s-allocate}). 


\begin{algorithm}
\caption{2S-RSA algorithm}\label{alg:2s-rsa}

\begin{small}

\begin{flushleft}
        \textbf{INPUT:} network topology, potential attack scenarios, bunkers' locations, demands $d \in D$ to be allocated, parameter $|P|$ \\ 
        \textbf{OUTPUT:} allocation rules for all demands
\end{flushleft}

\begin{algorithmic}[1]

\STATE  calculate $VUL\_M$ \label{alg:2s-find-matrix}
\STATE  sort $d \in D$ in decreasing $h_d$ \label{alg:2s-sort}
\FOR{\textbf{each} $d \in D$}
\STATE $P_d \gets \emptyset $
\ENDFOR

\tcp{\textbf{Routing assignment}}
\FOR{\textbf{each} $d \in D$} \label{alg:2s-ra-start}
\FOR{$p \in P$}
\STATE $p \gets$ \textit{FindNextPath}(..., $VUL\_M$, $P_d$)\label{alg:2s-findpath}
\STATE $P_d = P_d \cup p$
\ENDFOR
\ENDFOR \label{alg:2s-ra-end}

\tcp{\textbf{Spectrum assignment}}
\FOR{\textbf{each} $d \in D$} \label{alg:2s-sa-start}
\FOR{$p \in P_d$}
\STATE $c \gets$ \textit{FindChannel}(..., \textit{p})\label{alg:2s-findchannel}
\STATE allocate demand \textit{d} using path \textit{p} and channel \textit{c}\label{alg:2s-allocate}
\ENDFOR
\ENDFOR \label{alg:2s-sa-end}

\end{algorithmic}
\end{small}
\end{algorithm}

\begin{algorithm}
\caption{\textit{FindNextPath()} function}\label{alg:findnextpath}

\begin{small}

\begin{flushleft}
        \textbf{INPUT:} network topology, potential attack scenarios, bunkers' locations, network vulnerability matrix $VUL\_M$, demand \textit{d}, already selected paths $p \in P_d$ for demand \textit{d}  \\ 
        \textbf{OUTPUT:} selected routing path for demand \textit{d}
\end{flushleft}

\begin{algorithmic}[1]

\STATE $p \gets$ $Find\_SLD\_Path$(..., $P_d$, $VUL\_M$)\label{alg:fnp_find_sld_path}
\IF{$p \neq \emptyset$}
\RETURN \textit{p}
\ELSE
\STATE $VUL\_M^\prime$ $\gets$ $FindModified\_VUL\_M$($P_d$, $VUL\_M$)\label{alg:fnp_find_modified_matrix}
\STATE $p \gets$ \textit{FindShortestPath}(..., $VUL\_M^\prime$)\label{alg:fnp_find_spath_in_modified}
\RETURN \textit{p}
\ENDIF

\end{algorithmic}
\end{small}
\end{algorithm}

Note that 2S-RSA solves first the routing assignment problem and therefore it prioritizes network resilience criterion over spectrum usage. The algorithm's process does not depend on the values of $c_{spec}, c_{res}$ coefficients. 

\subsection{RSA reference methods}\label{sec:ref-algs}

Two reference methods are FF-RSA and LD-RSA \cite{Christodoulopoulos_lightwave11, Klinkowski_osn18}. FF-RSA sorts demands in decreasing value of their bitrate and then allocates one by one using shortest possible paths and first-fit channels. LD-RSA applies the same ordering policy. Then, for each demand it finds 40 shortest paths and saves them as candidate paths. The first of these paths is saved as a~first selected path and removed from the set of candidate paths. Then, LD-RSA runs $|P|-1$ times a~special procedure in order to select next routing paths (from the set of candidate paths) and prioritize link-disjoitness. Let a~similarity metric of two paths be equal to the number of their common links. To select \textit{i}-th routing path (where $i > 1$), the algorithm calculates a~cost for each candidate path. To this end, LD-RSA calculates and sums similarity metrics with all already selected paths. The lowest-cost candidate path is then selected for allocation and removed from the set of candidate paths.

\section{Results}\label{sec:experiments}

This section presents results of extensive numerical experiments, which are divided into three parts: tuning of the 1S-RSA algorithm, comparison of heuristic methods dedicated to solve BLRSA problem and case study focused on the efficient design of EMP-resilient elastic optical network. 

\subsection{Simulation setup}

\begin{figure*}[!t]
\centering 

\begin{subfigure}[b]{0.25\textwidth} \includegraphics[width=\textwidth]{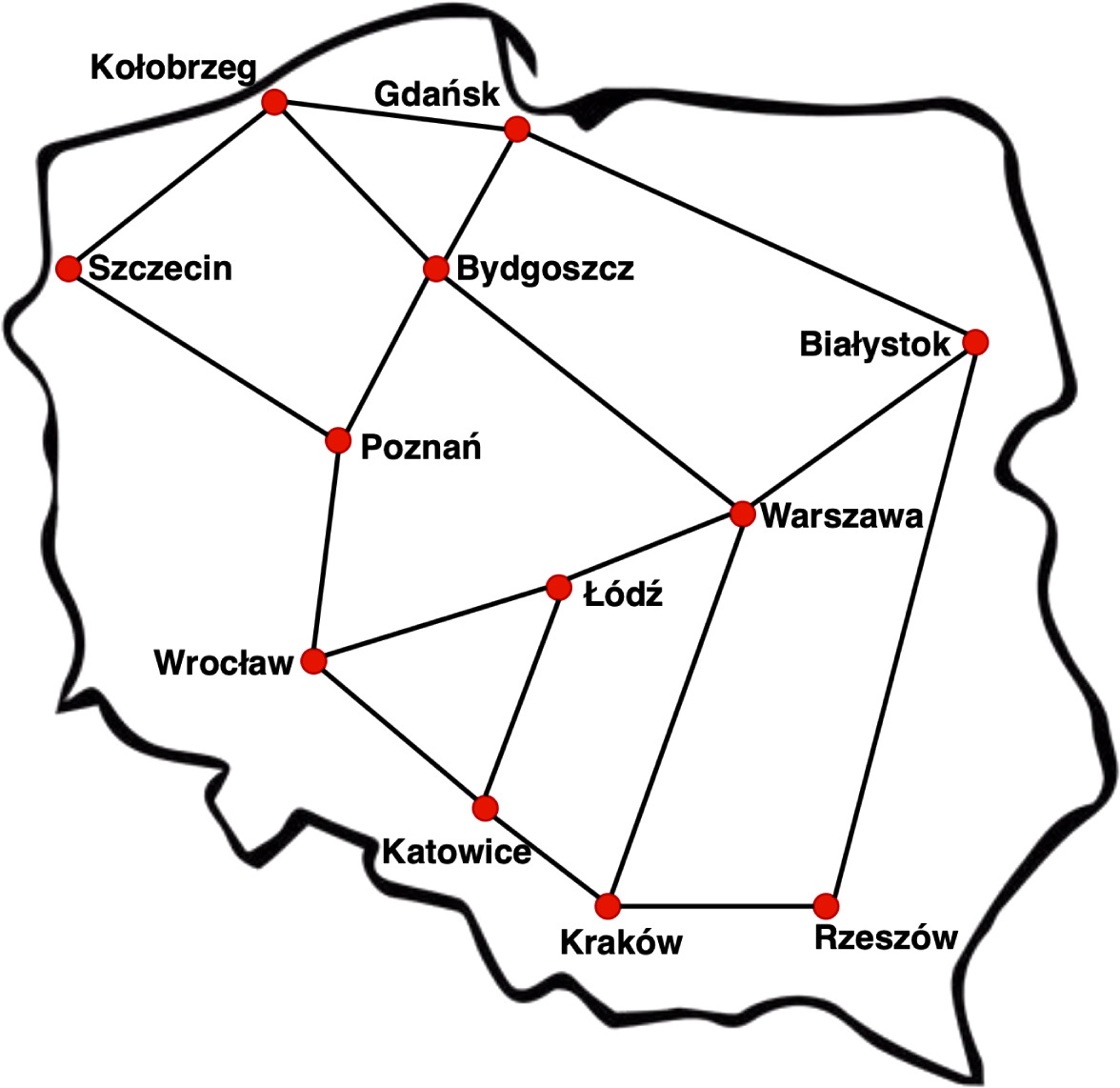} \caption{PL12} \label{fig:pl12} \end{subfigure}
\begin{subfigure}[b]{0.25\textwidth} \includegraphics[width=\textwidth]{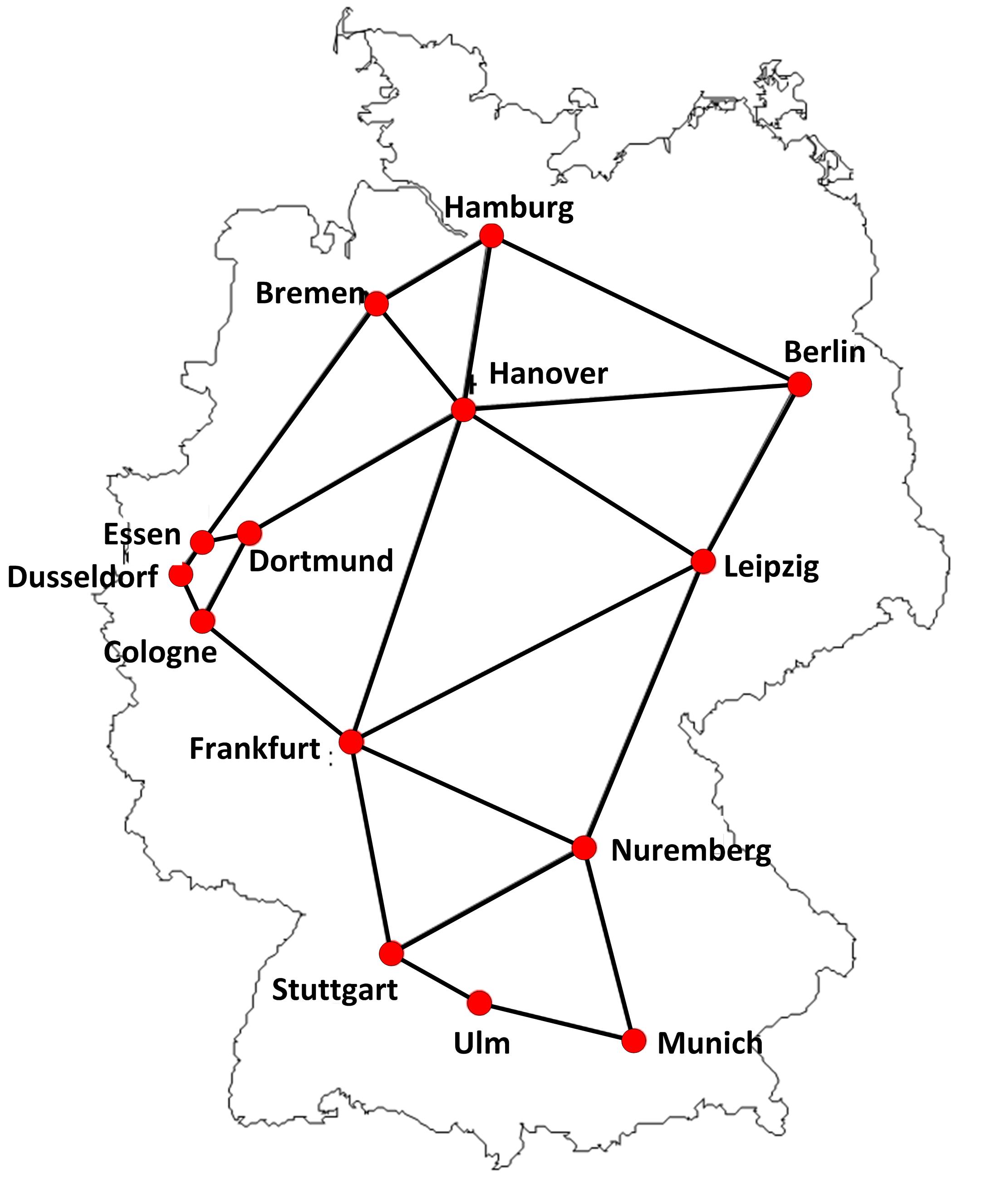} \caption{DT14} \label{fig:dt14} \end{subfigure}
\begin{subfigure}[b]{0.32\textwidth} \includegraphics[width=\textwidth]{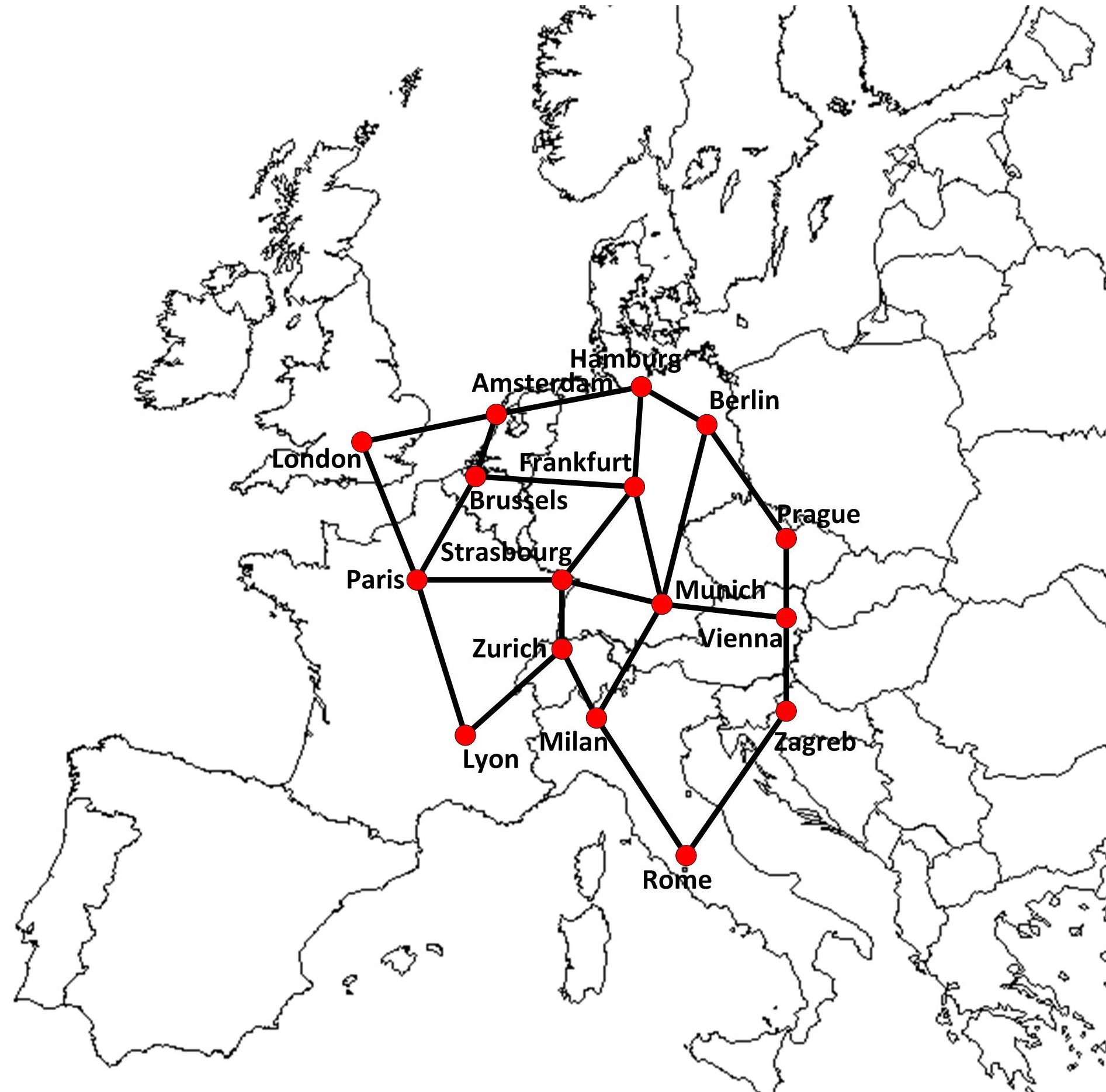} \caption{Euro16} \label{fig:euro16} \end{subfigure}

\caption{Network topologies used in the experiments}\label{fig:topologies} 
\end{figure*}

We use three realistic network topologies: PL12 (12 nodes, 36 links, 199 km average link length), DT14 (14 nodes, 46 links, 186 km average link length) and Euro16 (16 nodes, 48 links, 322 km average link length). The first two topologies are models of national networks of, respectively, Poland and Germany while the last one is a~model of European core network. The topologies are depicted in Fig.~\ref{fig:topologies}. 

For each of the topologies, we define two groups of testing scenarios -- \textit{small} and \textit{large}. The first group consists of unrealistic small problem instances for which optimal results can be found. Hence, it is used only for algorithms' comparison. The latter group presents realistic test cases which are simultaneously too complex to be solved optimally in a~reasonable time. However, these scenarios allow to present practical case study. 

Small group of scenarios consists of 5 different demand sets with overall traffic volume equal to 1 Tbps. The demand end nodes are selected uniformly at random while their bitrates are randomly selected from the range of 50--400 Gbps. Each demand set is considered with one set of potential attack scenarios. It assumes one potential attack on each network node (i.e., $|A|=|V|$) with destructive range limited to the attack target. The values of jamming ranges, concurrently, are selected uniformly at random from the range 10 -- \textit{X} km, wherein we consider $X = 50, 75, 100$ km. In turn, we have 10 different test cases, whose can be considered assuming various values of parameters $|P|$, $|B|$, $c_{spec}$ and $c_{res}$. If not stated otherwise, all presented results for small scenarios are averaged over 10 test cases (separately for each value of $|P|$, $|B|$, $c_{spec}$ and $c_{res}$ parameters). 

For the large group of scenarios we define 30 different demand sets with overall traffic volume equal to 40 Tbps and 30 different sets of identified potential attack scenarios. The demand end nodes are selected uniformly at random while their bitrates are randomly selected from the range of 50--500 Gbps. Next, each of the sets of identified attacks includes definition (i.e., attack target, destructive and jamming range) of $|A|=3 \cdot |V|$ attacks. The targets of the attacks are selected uniformly at random while all destructive ranges are limited to the target ones. The values of jamming ranges are selected uniformly at random from the range 10 -- \textit{X} km, wherein we consider $X = 50, 100, 150, ..., 550, 600$ km. For a~particular \textit{X} value, we have 30 demands sets and 30 sets of identified attack scenarios. Therefore, we have 900 different test cases. Moreover, each of these test cases can be considered assuming various values of parameters $|P|$, $|B|$, $c_{spec}$ and $c_{res}$. If not stated otherwise, all presented results for large scenarios are averaged over 900 test cases (separately for each value of $|P|$, $|B|$, $c_{spec}$ and $c_{res}$ parameters). 

In the experiments, we analyze three optimization criteria: average lost flow (LF) per attack measured in Gbps (representing network resilience to EMP attacks), maximum slice (MS) index (denoting spectrum usage) and weighted objective function (given by eq. (\ref{node_design_obj})) that combines both former criteria. 

Due to the limited size of this paper, in all parts of the experiments we present results only for selected configurations, which present the general observed trends.

Regarding EON physical realization, we work under the assumption that network nodes are equipped with coherent transceivers that operate at a~fixed baudrate, where each transceiver transmits/receives an optical signal that occupies 3 frequency slices (i.e., 37.5 GHz). To calculate the number of slices required to realize a~demand on a~routing path, we use network physical model proposed in \cite{Khodashenas_lighttech16}. Four modulations are available (i.e., BPSK, QPSK, 16-QAM, 32-QAM) wherein always the most spectrally-efficient one is selected (which simultaneously supports transmission on the given path length). Table~\ref{tab:modulations} presents supported bitrate and transmission distance for each modulation. To separate neighbouring connections, we use 12.5~GHz guard-band. 

\subsection{Tuning of 1S-RSA algorithm}

\begin{figure}[t]
\centering 

\begin{subfigure}[b]{0.49\textwidth} \includegraphics[width=\textwidth]{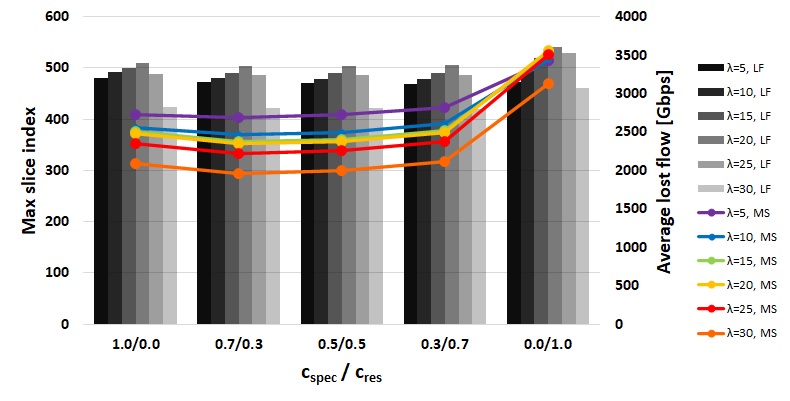} \caption{PL12, bunkers' policy: NodalDegree} \label{fig:pl12_tuning_x1} \end{subfigure}
\begin{subfigure}[b]{0.49\textwidth} \includegraphics[width=\textwidth]{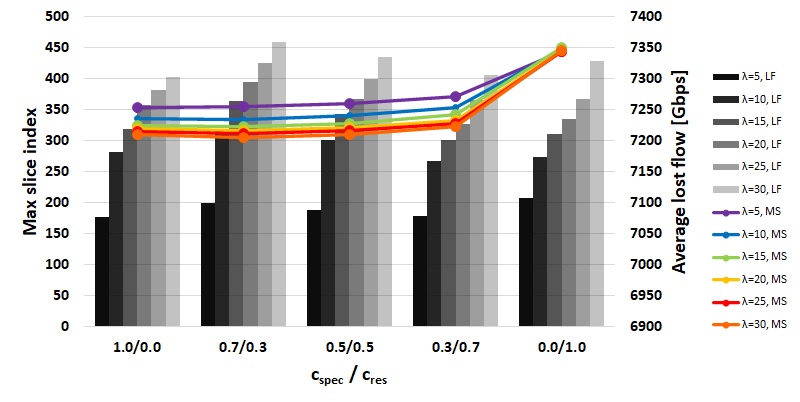} \caption{DT14, bunkers' policy: AvgNeighbour} \label{fig:dt14_tuning_x1} \end{subfigure}

\caption{Efficiency of 1S-RSA as a~function of $\lambda$  value: average lost flow per attack (bars) and maximum slice index (lines) for $X=200km, |P|=3, |B|=3$}\label{fig:tuning_x1} 
\end{figure}

\begin{figure}[t]
\centering 

\includegraphics[width=0.49\textwidth]{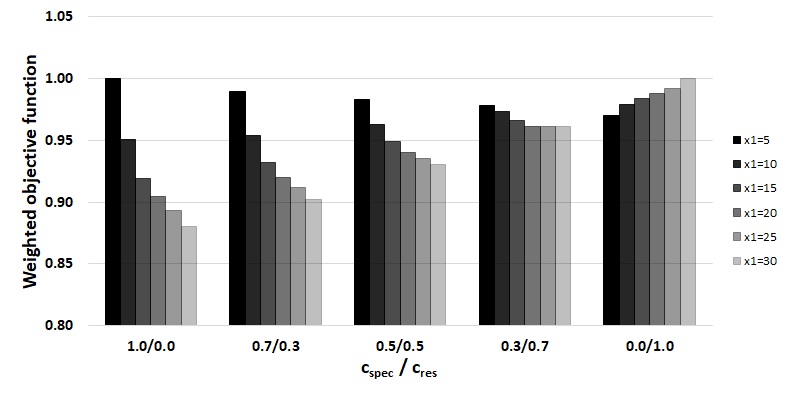} \label{fig:euro16_tuning_x1}

\caption{Efficiency of 1S-RSA as a~function of $\lambda$  value: weighted objective function for DT14, AvgNeighbour bunkers' policy, $X=200km$, $|P|=3$, $|B|=3$}\label{fig:tuning_x1_wmetric} 
\end{figure}

The first part of the experiments is tuning of 1S-RSA method, i.e., selection of beneficial $\lambda$ value. The tuning is performed separately for each network topology and each bunkers' location policy. Based on our previous experience with RSA optimization, we consider $\lambda$ $\in \{ 5, 10, 15, 20, 25, 30\}$. We make use of large set of testing scenarios and perform analysis assuming $c_{spec}/c_{res} = 1.0/0.0, 0.9/0.1, ..., 0.0/1.0$. In Fig.~\ref{fig:tuning_x1} we present averaged results in terms of two optimization criteria -- average lost flow (per attack) and spectrum width required to support all demands. For PL12 topology (see Fig.~\ref{fig:pl12_tuning_x1}), 1-RSA yields best results in terms of both criteria with $\lambda=30$ for all studied $c_{spec}/c_{res}$ combinations. For DT14 and Euro16 the situation differs. We observe the decreasing spectrum usage and increasing lost flow with increasing $\lambda$  value. The selection of $\lambda$  value is therefore a~trade-off between the both criteria. To face that problem and determine recommended $\lambda$  value, we use the weighted objective function given by formula (\ref{node_design_obj}). In the calculations we assume MAX\_SPEC to be the~maximum spectrum usage obtained for any $\lambda$  value and MAX\_LOSS to be the~maximum lost flow obtained for any $\lambda$. Fig.~\ref{fig:tuning_x1_wmetric} presents results of the tuning process in terms of weighted objective function for the same configuration as Fig.~\ref{fig:dt14_tuning_x1}. For further experiments, we recommend $\lambda$  values minimizing that function. The tuning results (i.e., recommended $\lambda$  values) are summarized in Table~\ref{tab:x1_tuning}.

\begin{table*}[!h]
    \centering
    \caption{Results of 1S-RSA tuning -- recommended values of $\lambda$  parameter accordingly for PL12/DT14/Euro16}
    \label{tab:x1_tuning}
    \begin{footnotesize}
	\begin{tabular}{|c|c|c|c|c|c|} \hline 

\multirow{2}{*}{\textbf{$c_{spec}/c_{res}$}} & \multicolumn{5}{c|}{\textbf{Bunkers' location policy}} \\ \cline{2-6}
{ } & \textbf{Adaptive/Avg} & \textbf{Adaptive/Max }& \textbf{AvgNeighbour} & \textbf{MinNeighbour} & \textbf{NodalDegree} \\ \hline \hline 
 \textbf{1.0/0.0} &   30/30/30 & 30/30/30 &  30/30/30 & 30/30/30 & 30/30/30 \\ \hline
 \textbf{0.9/0.1} &   30/30/25 & 30/30/25 &  30/30/25 & 30/30/25 & 30/30/25 \\ \hline
 \textbf{0.8/0.2} &   30/30/25 & 30/30/25 &  30/30/25 & 30/30/25 & 30/30/25 \\ \hline
 \textbf{0.7/0.3} &   30/30/25 & 30/30/25 &  30/30/25 & 30/30/25 & 30/30/25 \\ \hline
 \textbf{0.6/0.4} &   30/30/15  & 30/30/25  &  30/30/15 & 30/30/25  & 30/30/25 \\ \hline
 \textbf{0.5/0.5} &   30/30/5  & 30/30/5  &  30/30/5  & 30/30/5  & 30/30/15  \\ \hline
 \textbf{0.4/0.6} &   30/30/5  & 30/30/5  &  30/30/5  & 30/30/5  & 30/30/5  \\ \hline
 \textbf{0.3/0.7} &   30/30/5  & 30/20/5  &  30/30/5  & 30/30/5  & 30/25/5  \\ \hline
 \textbf{0.2/0.8} &   30/30/5  & 30/5/5  &  30/20/5  & 30/5/5  & 30/20/5  \\ \hline
 \textbf{0.1/0.9} &   30/5/5  & 30/5/5  &  30/5/5  & 30/5/5  & 30/5/5  \\ \hline
 \textbf{0.0/1.0} &   30/5/5   & 30/5/5   &  30/5/5   & 30/5/5   & 30/5/5   \\ \hline
   
    \end{tabular}
   \end{footnotesize}
\end{table*}

\subsection{Comparison of algorithms}\label{sec:alg_cmp}

In this section, we focus on the comparison of the heuristic algorithms dedicated to solve BLRSA problem. Four heuristic methods are compared. Two of them are completely novel proposals (i.e., 1S-RSA and 2S-RSA) and two approaches were proposed for other RSA formulations and adapted to solve the BLRSA (i.e., FF-RSA and LD-RSA). The comparison study is divided into two parts depending on the applied set of test cases -- small or large. In the first part, we use small test cases for which optimal results can be obtained by means of mathematical model (see Section~\ref{sec:blrsa-model}) implemented in IBM CPLEX Solver 12.1. Concurrently, for large test cases it is impossible to obtain optimal results in an acceptable time (up to several hours per test case) and using reasonable amount of computing resources (up to 20GB RAM memory). Therefore, only heuristic results are taken into consideration. 

\subsubsection{Small scenarios}

Table~\ref{tab:alg_cmp_small} presents algorithms comparison for small group of testing scenarios and selected configurations. Taking into account network resilience criterion, 2S-RSA is the best while 1S-RSA is second best method. However, the differences between them are irrelevant. In case of spectrum usage criterion, 2S-RSA performs the best. On the second place there are 1S-RSA and LD-RSA, whose represent similar performance for small scenarios. The processing time of all heuristic methods is very short and acceptable (less than a~second for a~single test case). The ILP calculations take much longer especially when optimization covers network resiliency. Based on the comparison for small scenarios, we expect that 1S-RSA and 2S-RSA might be two best methods, however the thesis should be tested for larger and more realistic scenarios. 

\begin{table}[!h]
    \centering
    \caption{Comparison of the algorithms for small scenarios, $|P|=1$ and  $|B|=2$}
    \label{tab:alg_cmp_small}
    \begin{footnotesize}
	\begin{tabular}{c|c|c|c|c|c} \\

 & \textbf{ILP} & \textbf{1S-RSA} & \textbf{2S-RSA} & \textbf{FF-RSA} & \textbf{LD-RSA}  \\ \hline \hline 
 
 \multicolumn{6}{c}{\textbf{average lost flow [Gbps] per attack} for $c_{spec}=0, c_{res}=1$} \\ \hline
 
$X=50km$ &	101.2 &	104.3 &	101.2 &	114.5 &	114.5 \\ \hline
$X=75km$ & 101.2 &	104.3 &	101.2 &	113.7 &	114.5 \\ \hline
$X=100km$ & 115.8 &	185.8 &	180.4 &	216.9 &	216.9 \\ \hline

 \multicolumn{6}{c}{\textbf{average max slice index} for $c_{spec}=1, c_{res}=0$} \\ \hline
 
$X=50km$ & 11.25	& 15.75	& 15	& 18	& 15.75 \\ \hline
$X=75km$ & 11.25	& 15.75	& 15	& 18	& 15.75 \\ \hline
$X=100km$ & 11.25	& 18	& 15.75	& 18	& 15.75 \\ \hline

 \multicolumn{6}{c}{\textbf{average processing time [s] }for $c_{spec}=1, c_{res}=0$} \\ \hline
 
$X=50km$ & 1.313 	& 0.054	& 0.002	& 0.001	& 0.006 \\ \hline
$X=75km$ & 1.254	& 0.045	& 0.001	& 0.001	& 0.006 \\ \hline
$X=100km$ & 1.371  & 0.046	& 0.002	& 0.001	& 0.006 \\ \hline

 \multicolumn{6}{c}{\textbf{average processing time [s] }for $c_{spec}=0, c_{res}=1$} \\ \hline
 
$X=50km$ & 334.316	& 0.054	& 0.002	& 0.001	& 0.006 \\ \hline
$X=75km$ & 335.802	& 0.045	& 0.001	& 0.001	& 0.006 \\ \hline
$X=100km$ & 336.218	& 0.046	& 0.002	& 0.001	& 0.006 \\ \hline

    \end{tabular}
     \end{footnotesize}
\end{table}

\subsubsection{Large scenarios}

\begin{figure}[!h]
\centering 

\begin{subfigure}[b]{0.49\textwidth} \includegraphics[width=\textwidth]{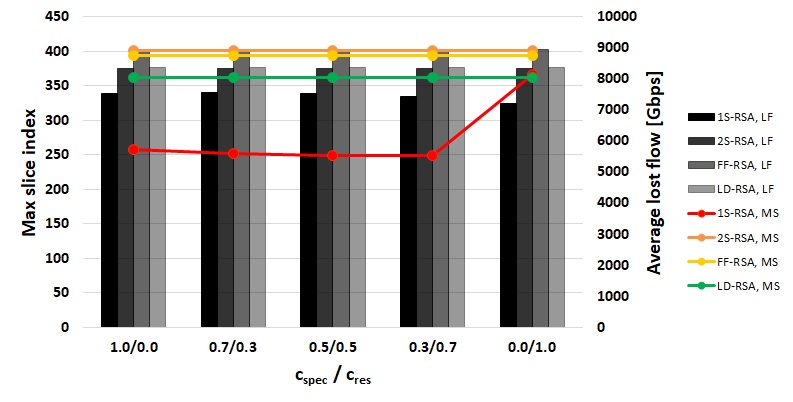} \caption{PL12, bunkers' policy: Adaptive/Avg} \label{fig:pl12_cmp_large} \end{subfigure}
\begin{subfigure}[b]{0.49\textwidth} \includegraphics[width=\textwidth]{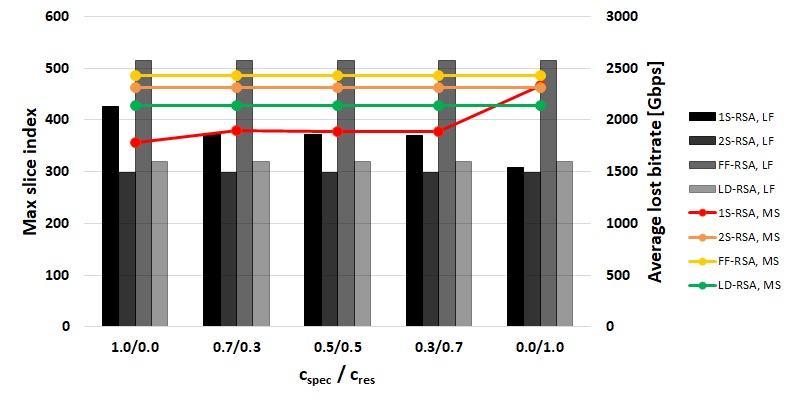} \caption{Euro16, bunkers' policy: Adaptive/Avg} \label{fig:euro16_cmp_large} \end{subfigure}

\caption{Comparison of algorithms for large scenarios: average lost flow per attack (bars) and maximum slice index (lines) for $|B|=3$, $|P|=3$}\label{fig:cmp_large} 
\end{figure}

\begin{figure}[!h]
\centering 

\includegraphics[width=0.49\textwidth]{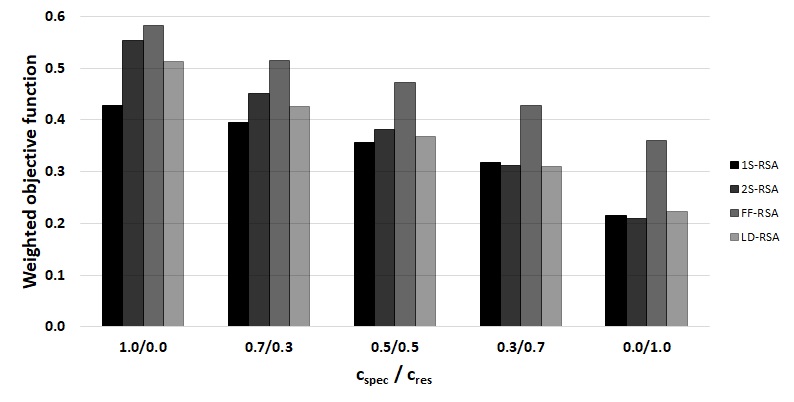} \label{fig:euro16_cmp_large_final}

\caption{Comparison of algorithms for large scenarios: weighted metric for Euro16, AvgNeighbour policy, $|B|=3$ and $|P|=3$}\label{fig:cmp_large_wmetric} 
\end{figure}

 Fig.~\ref{fig:cmp_large} presents algorithms comparison for large group of testing scenarios. Results are shown as~a~function of $c_{spec}/c_{res}$, however, only results of 1S-RSA differ with these coefficients (only that algorithm implements mechanisms to prioritize two objective criteria). In case of PL12 we observe that 1S-RSA significantly outperforms rest of the methods for all studied cases and taking into account both criteria. Thus, we recommend to employ that algorithm for calculations for PL12. In case of Euro16 (and DT14 that follows the same trends), results differ depending on the assumed criterion and coefficients $c_{spec}/c_{res}$. Considering lost flow criterion, 2S-RSA always ouperforms other methods, since it prioritizes (by addressing it first) resilience related criterion. We also notice increasing efficiency of 1S-RSA with increasing value of $c_{res}$. For $c_{res}=1$ 1-S-RSA protects almost the same amount of traffic as 2S-RSA. When analyzing spectrum usage criterion, we notice that for all cases where that criterion matters (i.e., $c_{spec} \neq 0$) 1S-RSA yields the best results. Based on the presented data we can state that 1S-RSA is best method for $c_{spec}=1$ while 2S-RSA is the most suitable for $c_{res}=1$. To determine best method for the rest of cases, we use weighted objective function (see eq.~\ref{node_design_obj}) assuming MAX\_SPEC to be highest spectrum usage obtained by any of the comparing algorithms and MAX\_LOSS is the highest lost flow value obtained by any of the comparing algorithms. Fig.~\ref{fig:cmp_large_wmetric} presents algorithms' comparison in terms of the weighted objective function for Euro16. 1S-RSA performs best for the majority of cases. Only for selected configurations with $0.9 \leq c_{res}$ 2S-RSA yields better results. Table~\ref{tab:alg_cmp_large} summarizes the comparison by recommending most efficient method for further experiments based on the analysis of weighted objective function. 

\begin{table}[!h]
    \centering
    \caption{Comparison of algorithms for large scenarios -- best algorithms according to weighted objective function}
    \label{tab:alg_cmp_large}
    \begin{footnotesize}
	\begin{tabular}{|c|c|c|c|} \hline 

\multirow{2}{*}{\textbf{$c_{spec}/c_{res}$}} & \multicolumn{3}{c|}{\textbf{Network topology}} \\ \cline{2-4}
{ } & \textbf{PL12} & \textbf{DT14}& \textbf{Euro16} \\ \hline \hline 
 \textbf{1.0/0.0} &   1S-RSA & 1S-RSA & 1S-RSA \\ \hline
 \textbf{0.9/0.1} &   1S-RSA & 1S-RSA & 1S-RSA \\ \hline
 \textbf{0.8/0.2} &   1S-RSA & 1S-RSA & 1S-RSA \\ \hline
 \textbf{0.7/0.3} &   1S-RSA & 1S-RSA & 1S-RSA \\ \hline
 \textbf{0.6/0.4} &   1S-RSA  & 1S-RSA  & 1S-RSA \\ \hline
 \textbf{0.5/0.5} &   1S-RSA  & 1S-RSA  & 1S-RSA  \\ \hline
 \textbf{0.4/0.6} &   1S-RSA  & 1S-RSA  & 1S-RSA  \\ \hline
 \textbf{0.3/0.7} &   1S-RSA  & 1S-RSA  & 1S-RSA  \\ \hline
 \textbf{0.2/0.8} &   1S-RSA  & 1S-RSA  & 1S-RSA  \\ \hline
 \textbf{0.1/0.9} &   1S-RSA  & 2S-RSA  & 1S-RSA  \\ \hline
 \textbf{0.0/1.0} &   1S-RSA   & 1S-RSA & 2S-RSA  \\ \hline
   
    \end{tabular}
    \end{footnotesize}
\end{table}

Eventually, in the Table~\ref{tab:alg_cmp_time} we compare average processing time of all of the algorithms. All methods are characterized by short and acceptable processing time -- calculation takes up to 0.4 sec for a~single test scenario.

\begin{table}[!h]
    \centering
    \caption{Comparison of algorithms for large testing scenarios -- average processing time [s] }
    \label{tab:alg_cmp_time}
    \begin{footnotesize}
	\begin{tabular}{|c|c|c|c|c|}\hline

{ } & \textbf{1S-RSA} & \textbf{2S-RSA} & \textbf{FF-RSA} & \textbf{LD-RSA} \\ \hline \hline 
 \textbf{PL12}      & 0.392 &   0.010 &     0.009 &     0.080 \\ \hline
 \textbf{DT14}      & 0.345 &   0.012 &     0.010 &     0.105 \\ \hline
 \textbf{Euro16}    & 0.109 &   0.014 &     0.012 &     0.144 \\ \hline

    \end{tabular}
    \end{footnotesize}
\end{table}

\subsection{Comparison of bunkers' location policies}\label{sec:bl_policies_cmp}

Here, we focus on the comparison of five bunkers' location policies. To this end, we use group of large testing scenarios and two best algorithms (according to Section~\ref{sec:alg_cmp}) -- 1S-RSA and 2S-RSA. We present results for $|P|=2$, $|B|=4$, \textit{X} = 200 km for PL12/DT14 and \textit{X} = 400 km for Euro16.

\begin{figure}[!h]
\centering 

\begin{subfigure}[b]{0.49\textwidth} \includegraphics[width=\textwidth]{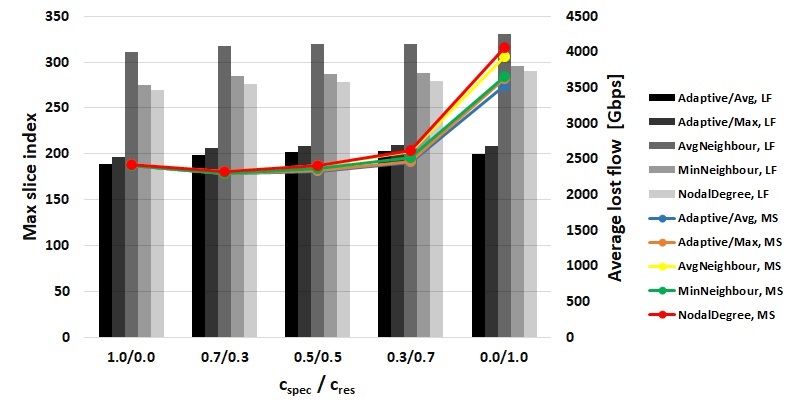} \caption{PL12, \textit{X}=200km} \label{fig:pl12_bl_cmp} \end{subfigure}
\begin{subfigure}[b]{0.49\textwidth} \includegraphics[width=\textwidth]{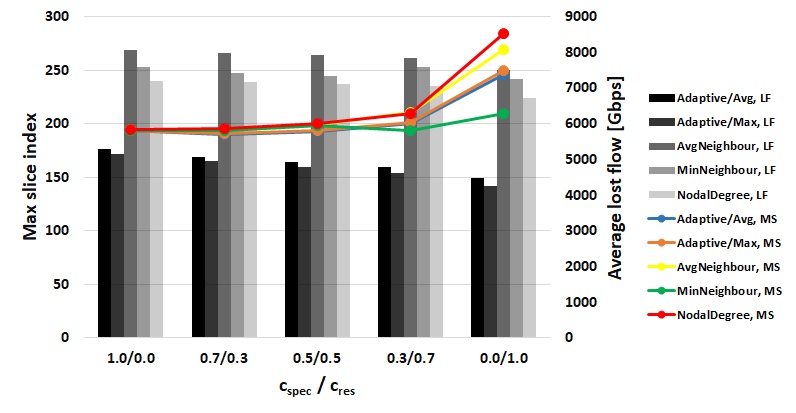} \caption{DT14, \textit{X}=200km} \label{fig:dt14_blcmp} \end{subfigure}
\begin{subfigure}[b]{0.49\textwidth} \includegraphics[width=\textwidth]{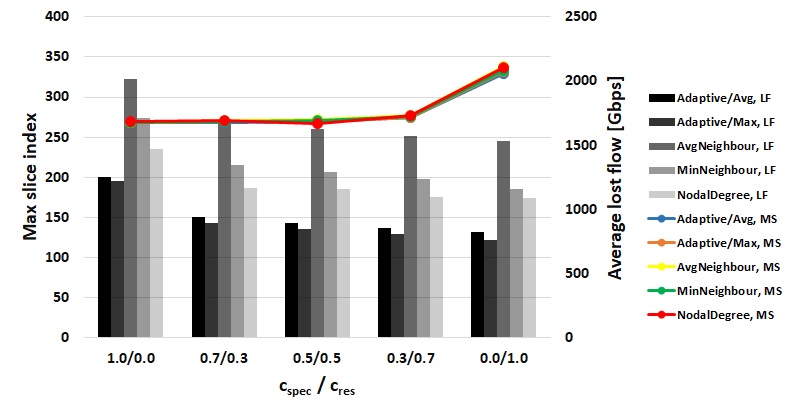} \caption{Euro16, \textit{X}=400km} \label{fig:euro16_blcmp} \end{subfigure}

\caption{Comparison of various bunkers' location policies: average lost flow per attack (bars) and maximum slice index (lines) for 1S-RSA with $|B| = 2$, $|P|=2$}\label{fig:bl_cmp} 
\end{figure}

Fig.~\ref{fig:bl_cmp} reports comparison for 1S-RSA presented as a~function of $c_{spec}/c_{res}$. Considering average lost flow criterion, two policies are outstanding -- the Adaptive/Avg and the Adaptive/Max. They bring best results for all studied $c_{spec}/c_{res}$ configurations. However, Adaptive/Avg is better for PL12 while Adaptive/Max is more suitable for DT14/Euro16. Taking into account spectrum usage criterion the significant differences between policies are observed only for PL12 and DT14. For PL12 best results are yielded for Adaptive/Avg. For DT14, they are observed for MinNeighbour policy (which does not perform satisfactory for the second criterion) while two Adaptive policies are second best. Based on the results for 1S-RSA, we reveal Adaptive/Avg to be the most efficient policy for PL12 and Adaptive/Max for DT14/Euro16. We also recommend that configuration for further experiments. 

\begin{figure}[!h]
\centering 

\begin{subfigure}[b]{0.49\textwidth} \includegraphics[width=\textwidth]{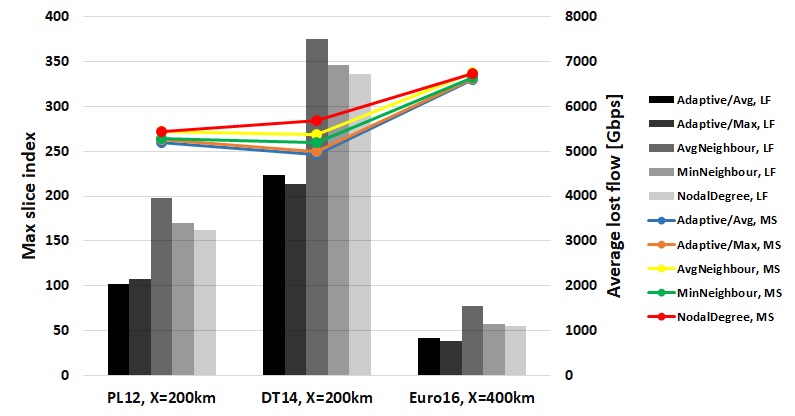} \end{subfigure}

\caption{Comparison of various bunkers' location policies: average lost flow per attack (bars) and maximum slice index (lines) for 2S-RSA with $|B| = 2$, $|P|=2$}\label{fig:bl_cmp_2step} 
\end{figure}

Fig.~\ref{fig:bl_cmp_2step} presents comparison for 2S-RSA. The results follow very similar trends to these from 1S-RSA and also prove high efficiency of Adaptive policies. 

\subsection{EMP-resilient network design -- case study}

In the last part of experiments we focus on the realistic case study. We make use of the proposed optmization approaches and study benefits of the considered protection scheme and costs of the EMP-resiliency provisioning. We also examine vulnerabilities of three realistic network topologies to nuclear weapon/EMP attacks and identify their critical nodes (taking into account that type of threat). 

\subsubsection{Protection efficiency as a~function of the number of protecting paths and bunkers}

In the first part of the case study, we evaluate influence of the number of available bunkers and number of applied routing paths on the network performance in terms of its resilience and the spectrum usage. We use group of large testing scenarios assuming $|B| = 0, 1, 2, ..., 8$ and $|P| = 1, 2, 3, 4$. 

\begin{figure}[!h]
\centering 
\includegraphics[width=0.49\textwidth]{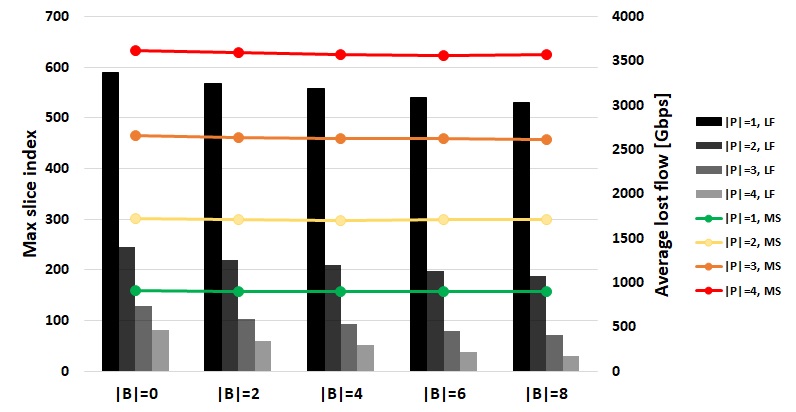} 
\caption{Average lost flow per attack (bars) and maximum slice index (lines) as a~function of $|B|$ and $|P|$ for PL12, $X=200km$, $c_{spec}=1$ and $c_{res}=0$}\label{fig:b_p_cmp}
\end{figure}

Fig.~\ref{fig:b_p_cmp} presents network performance (measured using spectrum usage and average lost flow per attack) as a~function of parameters $|P|$ and $|B|$. Recall that $|P|$ denotes number of routing paths assigned for each demand while $|B|$ indicates the number of located bunkers in the network. As we can observe, the number of the applied paths influences both performance metrics, while the number of bunkers impacts in a~noticeable way only the average lost flow value. Considering parameter $|P|$, we observe a~decrease of both performance metrics with its increasing value. In case of $|B|$, we notice decrease of the lost flow value when the number of bunkers is higher. 

The application of multipath routing brings higher reduction of the average lost flow compared to the bunkers' implementation. In order to measure the benefits provided by these two elements of our protection approach, i.e., multipath routing ($|P|>1$) and bunkers implementation ($|B|>0$), we present in Tables~\ref{tab:gains_p} and \ref{tab:gains_b} gains of the application of these elements. In more detail, Table~\ref{tab:gains_p} reports how much of the lost flow can be saved by applying $|P|>1$ instead of $|P|=1$. Similarly, Table~\ref{tab:gains_b} presents how much of the lost traffic can be saved by applying $|B|>0$ instead of $|B|=0$. According to Table~\ref{tab:gains_p}, application of two routing paths for each demand and no bunkers implementation (i.e., $|P|=2$, $|B|=0$) allow to save 36\% of the traffic lost in case of single-path routing. We can additionally increase the saved flow value by implementing some bunkers in the network nodes. For instance, the additional implementation of only two bunkers (i.e., $|P|=2$, $|B|=2$) increases the amount of saved flow up to 49.9\% while six bunkers (i.e., $|P| = 2$, $|B| = \frac{|V|}{2} = 6$) brings savings of about 76.8\%. The application of more routing paths per demand allows to reach further savings. Then, according to Table~\ref{tab:gains_b}, application of two bunkers (i.e., $|B|=2$) allows to save 14.4\% of the traffic lost in the case of no bunkers implementation when $|P|=1$ is applied and up to 28\% when $|P|=4$ is used. 

\begin{table}[!h]
    \centering
    \caption{Gains provided by application of $|P|>1$ instead of $|P|=1$ for PL12, \textit{X} = 200km, $c_{spec}=0$ and $c_{res}=1$}
    \label{tab:gains_p}
    \begin{footnotesize}
	\begin{tabular}{c||ccc} \
& $|P|$=2 & $|P|$=3 & $|P|$=4  \\ \hline \hline
$|B|$=0 & 36.0\% & 47.0\% & 53.4\%  \\
$|B|$=2 & 49.9\% & 60.8\% & 66.8\% \\
$|B|$=4 & 60.4\% & 70.6\% & 75.9\% \\
$|B|$=6 & 68.1\% & 78.4\% & 83.3\% \\
$|B|$=8 & 76.8\% & 86.9\% & 90.8\% \\
\end{tabular}
\end{footnotesize}
\end{table}

\begin{table}[!h]
    \centering
    \caption{Gains provided by application of $|B|>0$ instead of $|B|=0$ for PL12, \textit{X} = 200km, $c_{spec}=0$ and $c_{res}=1$}
    \label{tab:gains_b}
    \begin{footnotesize}
	\begin{tabular}{c||cccc} 
& $|B|$=2 & $|B|$=4 & $|B|$=6 & $|B|=8$ \\ \hline \hline
$|P|=1$ & 14.4\% & 26.1\% & 35.2\% & 46.1\% \\
$|P|=2$ & 21.8\% & 38.2\% & 50.2\% & 63.7\% \\
$|P|=3$ & 26.0\% & 44.6\% & 59.2\% & 75.3\% \\
$|P|=4$ & 28.7\% & 48.2\% & 64.1\% & 80.2\% \\
\end{tabular}
\end{footnotesize}
\end{table}

\subsubsection{Cost of the proposed protection strategy}

The proposed protection strategy consists of two important elements -- bunkers location and mutlipath demands provisioning. Both of these elements provide very high network resilience, however, at extra cost. In case of bunkers implementation, the cost is determined by the financial cost of bunkers. In the case of multipath provisioning, it is mostly determined by the amount of required additional network resources. One of these resources is spectrum width. As presented in Fig.~\ref{fig:b_p_cmp}, required spectrum width increases meaningly with increasing number of applied paths per demand. In Table~\ref{tab:cost_spectrum} we detail this dependence and indicate how many times spectrum width required for $|P|>1$ is higher that the spectrum required for single-path allocation. The application of $|P|>1$ needs approximately $|P|$ times more spectrum compared to the single-path allocation. The bunkers implementation reduces a~little bit supplementary spectrum consumption. When the protection comes from two different mechanisms (i.e., multipath routing and bunkers), the allocation process can apply shorter and less geographically-distance paths, which can use more efficient modulation formats. 

\begin{table}[!h]
    \centering
    \caption{Cost of multipath routing: ratio of additional spectrum consumption for $|P|>1$ applied instead of $|P|=1 $ for PL12, \textit{X} = 200km, $c_{spec}=0$ and $c_{res}=1$}
    \label{tab:cost_spectrum}
    \begin{footnotesize}
	\begin{tabular}{c||ccc} 
	& $|P|=2$ & 	$|P|=3|$ & 	$|P|=4$ \\\hline \hline
$|B|=0$ &	 1.98 & 	 3.04 & 	 4.15 \\ 
$|B|=1$ &	 1.98 & 	 3.00 & 	 4.10 \\
$|B|=2$	&    1.92 &  	 2.93 & 	 3.99 \\ 
$|B|=3$ &	 1.92 & 	 2.94 & 	 3.96 \\
$|B|=4$ &	 1.91 & 	 2.97 & 	 4.02 \\
$|B|=5$ &	 1.91 & 	 2.96 & 	 4.00 \\
$|B|=6$ &	 1.92 & 	 2.96 & 	 4.02 \\
$|B|=7$ &	 1.91 & 	 2.93 & 	 4.02 \\
$|B|=8$ &	 1.95 & 	 2.93 & 	 4.03 \\

\end{tabular}
\end{footnotesize}
\end{table}

\subsubsection{Analysis of network vulnerability to nuclear weapon/EMP attacks of various jamming range}

\begin{figure}[!h]
\centering 

\begin{subfigure}[b]{0.49\textwidth} \includegraphics[width=\textwidth]{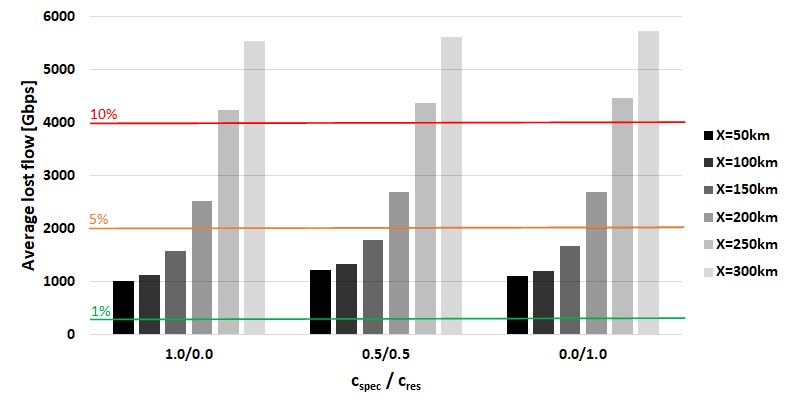}\caption{PL12}\end{subfigure}
\begin{subfigure}[b]{0.49\textwidth} \includegraphics[width=\textwidth]{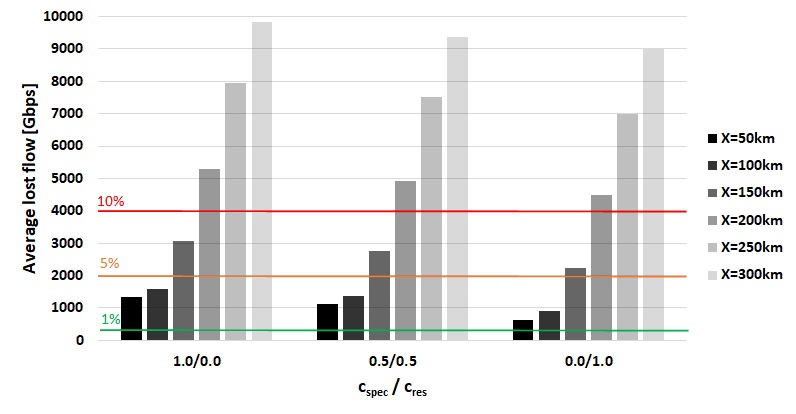}\caption{DT14}\end{subfigure}
\begin{subfigure}[b]{0.49\textwidth} \includegraphics[width=\textwidth]{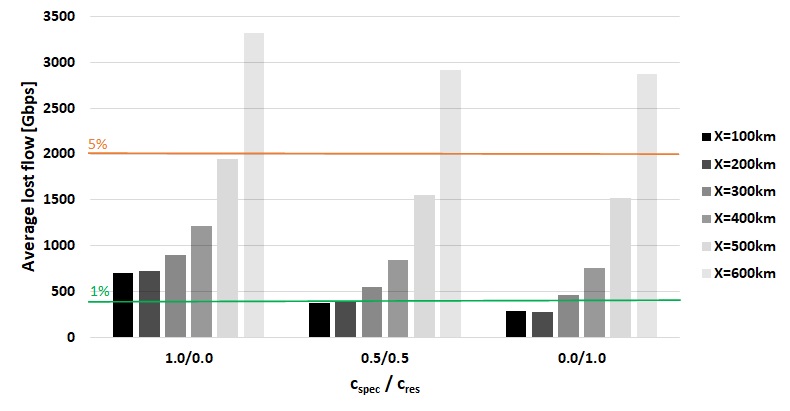}\caption{Euro16}\end{subfigure}

\caption{Average lost flow as a~function of attack jamming range, $c_{spec}$, $c_{res}$ for 1S-RSA with $|B|=4$, $|P|=2$}\label{fig:radius_cmp} 
\end{figure}

Eventually, we analyze vulnerability of three realistic network topologies to EMP attacks of different jamming range. Fig.~\ref{fig:radius_cmp} presents average lost flow per attack as a~function of jamming range and values of coefficients $c_{spec}/c_{res}$ for large testing scenarios. In order to better visualize lost traffic volume, we also mark three vertical lines referring to the specific ratio of overall traffic volume in the network -- green (indicating 1\% of overall traffic volume in the network), orange (indicating 5\%) and red (indicating 10\%).

The first obvious observation is fact that average lost flow increases with increasing attack jamming range, since more network elements are then affected. Mind that, the traffic loss does not increase identically for all topologies. We observe that generally DT14 suffers the most from EMP effects, then PL12 and eventually Euro16. In all studied cases, the average traffic loss exceeds 1\% (of overall traffic volume) for PL12 and DT14. For PL12, the efficient protection allows to reduce lost flow ratio to less than 5\% when $X \leq 150km$ and less than 7\% for $X \leq 200km$. In all other cases, the loss exceeds 10\%. In case of DT14, the loss reduction under 5\% is achieved for $X \leq 150km$, while reduction below 6-7\% for $X \leq 150km$. For the rest of cases, the loss is higher than 10\%. In case of Euro16, however, the efficient protection allows to reduce lost flow ratio to below or approximately 1\% for attacks with $X \leq 300km$, to 2\% for $X \leq 400km$, 4\% for $X \leq 500km$. Even for huge attacks with $X \leq 600km$, the loss flow value is smaller than 7\% of overall traffic volume. 

The obtained results prove high efficiency of the mechanisms of prioritization of network resilience criterion (i.e., application of high $c_{res}$ values) in 1S-RSA algorithm. The differences are clearly visible for DT14 and Euro16 topologies, where the application of $c_{res}=1$ instead of $c_{res}=0$ allows to save up to, respectively, 12.5\% (for $X=250km$) and 25\% (for $X=500$) of the traffic.  

\begin{figure*}[t]
\centering 

\begin{subfigure}[b]{0.25\textwidth} \includegraphics[width=\textwidth]{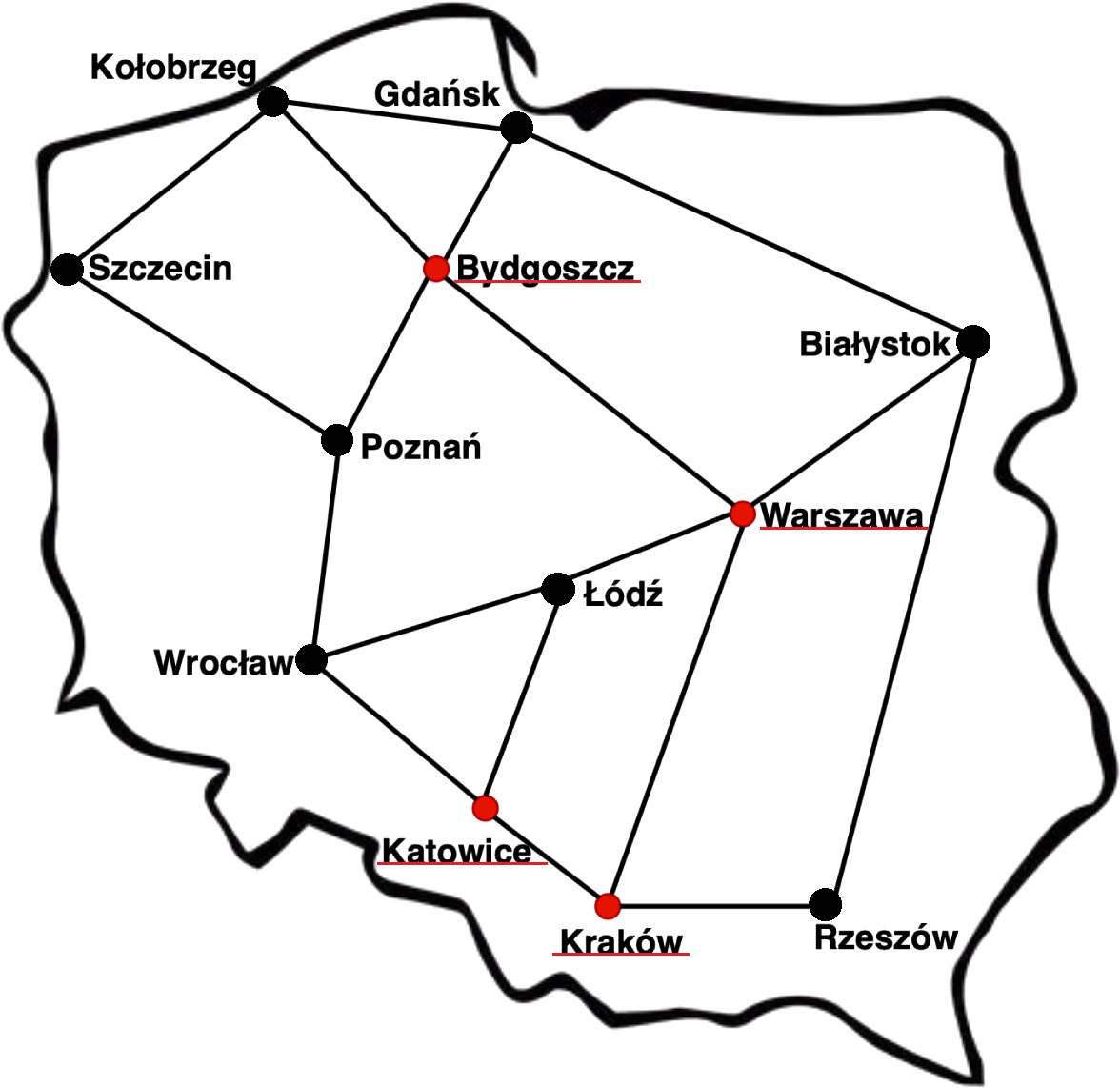} \caption{PL12} \label{fig:pl12_cnodes} \end{subfigure}
\begin{subfigure}[b]{0.25\textwidth} \includegraphics[width=\textwidth]{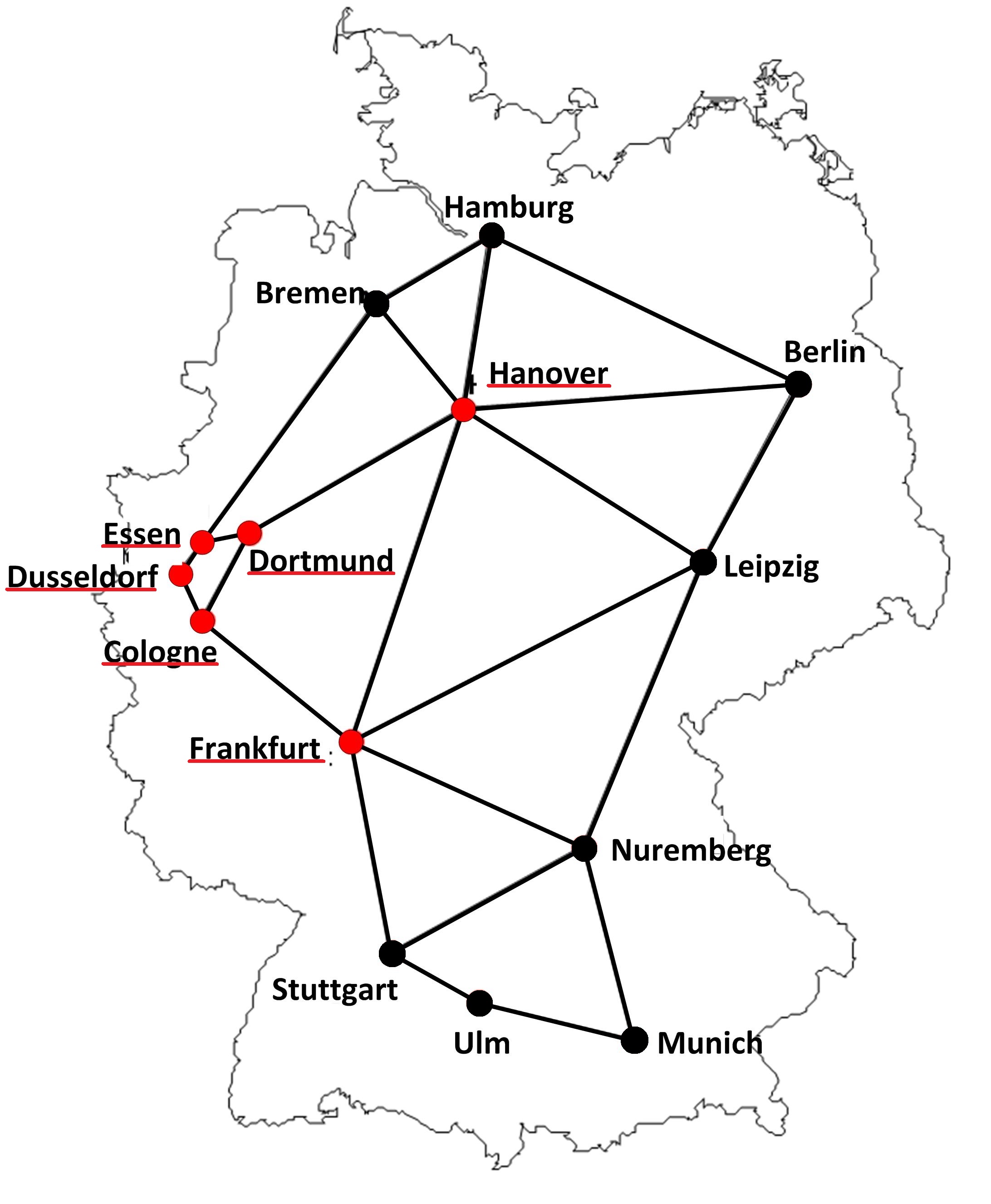} \caption{DT14} \label{fig:dt14_cnodes} \end{subfigure}
\begin{subfigure}[b]{0.32\textwidth} \includegraphics[width=\textwidth]{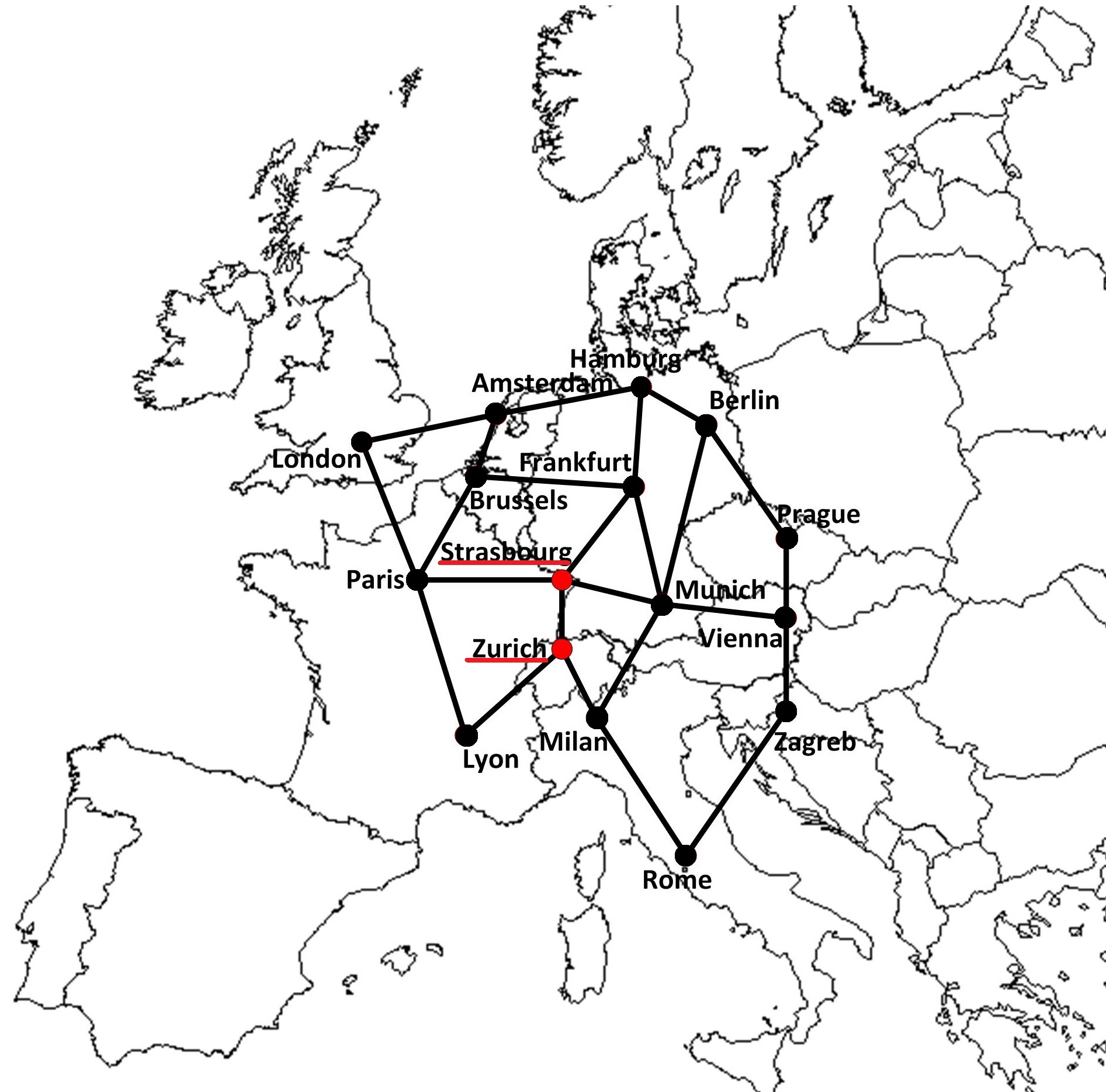} \caption{Euro16} \label{fig:euro16_cnodes} \end{subfigure}

\caption{Critical nodes (underlined and marked with red) of network topologies considering EMP-attack scenarios}\label{fig:critical_nodes} 
\end{figure*}

Let us now focus on the analysis of the examined network topologies in order to explain their different vulnerability to EMP-attacks. Table~\ref{tab:topologies_params} presents comparison of their topological characteristics including links lengths (min, max and average). The topology the most prone to harmful EMP effects (i.e., DT14) has the shortest links, while the topology the most resilient (i.e., Euro16) has the longest links. We can draw the important conclusion, that network vulnerability to EMP-attacks strongly depends on the lengths of network links. In more detail, the lengths should be compared to the attack jamming range. The links that are shorter might be problematic, since in case of an attack both their end nodes are simultaneously affected by the excessive EM radiation. Hence, nodes related to short links (i.e., links shorter than potential/expected attack jamming range) are critical nodes in terms of EMP attacks. Additionally, nodes with relatively high nodal degree might be candidates for critical nodes, since the failure of these nodes entails failure of many network links. Based on these two criteria, we have identified critical nodes for PL12, DT14 and Euro16. They are depicted in Fig.~\ref{fig:critical_nodes}.

Please note that each attack can be carefully planned, while each attacker might have different goals of the attack. For instance, the attacker may try to bypass all protection plans and mechanisms, so potentially negligible node would be affected. Therefore, the fully reliable network vulnerability analysis and protection against attacks is not possible. 

\begin{table}[!h]
    \centering
    \caption{Characteristics of the examined topologies}
    \label{tab:topologies_params}
    \begin{footnotesize}
	\begin{tabular}{c||ccc} 
& PL12 & DT14 & Euro16 \\ \hline \hline
no nodes                &   12 & 14 & 16 \\
no links                &   36 & 46 & 48 \\
avg nodal degree        &   2.9  & 3.3 & 3.0 \\
min link length [km]    &   70  & 37 & 147   \\
max link length [km]    &   360 & 353 & 517 \\
avg link length [km]    &   185 & 182 & 318 \\
\end{tabular}
\end{footnotesize}
\end{table}

\section{Conclusions}\label{sec:conclusions}

In this paper, we studied efficient design of optical network resilient to EMP attacks (or EMP following nuclear attack) optimizing simultaneously network resilience (measured as the average lost flow per potential attack) and spectrum usage. In order to increase network resiliency, we proposed dedicated protection scheme which implements multipath routing and bunkers implementation in selected network nodes. We formulated (using ILP technique) problem of bunkers location, routing and spectrum allocation in elastic optical networks. Since the problem is very challenging, we proposed two solution algorithm -- 1S-RSA and 2S-RSA. Next, we performed extensive numerical experiments using three realistic network topologies. The investigation was divided into three parts: (\textit{i}) tuning of the proposed approaches, (\textit{ii}) comparison of algorithms with respect to reference methods, (\textit{iii}) case study focused on the efficient design of network resilient to EMP attacks. In case study, we examined benefits and costs of the proposed protection scheme. We also analyzed the vulnerabilities of three realistic network topologies to EMP attacks and identified their critical nodes. The results of experiments prove high efficiency of the proposed protection scheme. It allows to save up to 90\% of traffic lost in the case of potential attack in an unprotected network. The efficiency of the proposed scheme is strongly determined by the number of applied routing paths per each demand and number of bunkers implemented in the network. Moreover, the case study shows that small local or national networks are prone to EMP attack effects due to relatively short link lengths. For that attack scenario, we identified three potential groups of critical nodes: (\textit{i}) nodes having geographically close neighbours, (\textit{ii}) nodes with high nodal degree, (\textit{iii}) nodes representing very important network nodes (i.e., nodes related to important cities, data centers, aggregated tremendous amount of users and data). 

In future work, we plan to further investigate bunkers location policies and BLRSA optimization using artificial intelligence and machine learning algorithms. 

\section*{Acknowledgment}
The work of R\'{o}\.{z}a~Go\'scie\'n was partially supported by the Foundation for Polish Science (FNP).

\bibliographystyle{elsarticle-num} 
\bibliography{goscien_hpm.bib}

\end{document}